# Development in times of hype: How freelancers explore Generative AI?

Mateusz Dolata
Department of Informatics
University of Zurich
Zurich, Switzerland
dolata@ifi.uzh.ch

Norbert Lange
Entschleunigung Norbert Lange
Kassel, Germany
nl@entschleunigung-lange.de

Gerhard Schwabe
Department of Informatics
University of Zurich
Zurich, Switzerland
schwabe@ifi.uzh.ch

## ABSTRACT

The rise of generative AI has led many companies to hire freelancers to harness its potential. However, this technology presents unique challenges to developers who have not previously engaged with it. Freelancers may find these challenges daunting due to the absence of organizational support and their reliance on positive client feedback. In a study involving 52 freelance developers, we identified multiple challenges associated with developing solutions based on generative AI. Freelancers often struggle with aspects they perceive as unique to generative AI such as unpredictability of its output, the occurrence of hallucinations, and the inconsistent effort required due to trial-and-error prompting cycles. Further, the limitations of specific frameworks, such as token limits and long response times, add to the complexity. Hype-related issues, such as inflated client expectations and a rapidly evolving technological ecosystem, further exacerbate the difficulties. To address these issues, we propose Software Engineering for Generative AI (SE4GenAI) and Hype-Induced Software Engineering (HypeSE) as areas where the software engineering community can provide effective guidance. This support is essential for freelancers working with generative AI and other emerging technologies.

## CCS CONCEPTS

• Software and its engineering → Software creation and management; • Computing methodologies → Artificial Intelligence.

## KEYWORDS

Generative AI, AI-based Systems, Challenges, Freelancers, Hype, SE for Generative AI, SE4GenAI, Hype-Induced SE, Hype-SE, Fashion, Product, Paradigm, Novelty, Qualitative Research

## 1 INTRODUCTION

Public interest in generative AI (GenAI) has recently surged. Numerous organizations are exploring the utility of large text-to-text and text-to-image foundation models, integrating them into their existing infrastructure or creating new applications. Small to mid-sized enterprises and startups, known for their agility, are prone to investigate new trends. Due to limited in-house development resources, they often depend on external freelancers who, contracted to deliver specific solutions, find themselves at the cutting edge of technological developments. Such freelancers take the role of explorers who open up new possibilities for their clients.

GenAI has limitations that may hinder independent developers working on applications. They urgently require guidance. The first step to providing effective support is to identify and understand the challenges they face when developing GenAI-based applications. Given GenAI's rapid uptake and the significant role freelance developers play in its adoption, we ask:

*RQ: What challenges do freelancers experience when developing solutions based on generative AI?*

We explore the perspectives of freelance developers on GenAI-related projects through the analysis of 52 interviews and survey responses. Approximately 60% of these freelancers work on small-scale projects as the sole developer, 20% participate in medium-sized projects with up to five team members, and the remaining 20% are either contractors in larger projects or operate a freelancing agency. These freelancers identify 99 challenges across 11 Software Engineering Body of Knowledge (SWEBOK) areas and 54 subcategories. These software engineering (SE) challenges arise from the inherent characteristics of GenAI as a technology and the associated hype. We use the term 'GenAI technology' to refer to the inherent features of GenAI as a computing *paradigm*, differentiating it from features of *products* that provide GenAI capabilities, such as large language models (LLMs) or foundation models and their APIs. The 'hype' refers to the *novelty* and *fashionableness* of GenAI among businesses. The characteristics of the paradigm, its products, its novelty, and its trendiness magnify known challenges and create new ones, as perceived by the freelancers.

Those findings are relevant to the SE community. Some of the issues identified by freelancers will likely impact the SE of GenAI-based applications for an extended period and, possibly, also beyond the context of freelancing. Further, future technologies will keep emerging as hype, requiring practitioners and researchers to manage hype-related phenomena. SE research must address these aspects. Thus, this paper suggests two research areas: (1) Software Engineering for Generative AI (SE4GenAI) as an extension of SE for AI (SE4AI) discourse already present in SE, and (2) Hype-Induced Software Engineering (HypeSE) to handle challenges associated with novel and fashionable technologies.

This manuscript contributes in three ways. First, it outlines challenges identified by freelancers who create GenAI-based solutions, adding a new, freelance-oriented perspective to SE4AI literature [15, 16, 37, 38, 42, 45, 57]. Second, it classifies these challenges highlighting the pivotal role of hype as a source of practical SE challenges. This complements earlier discussions of hype in SE, particularly relating to blockchain [14, 41, 43]. Third, it suggests new directions for SE research.

## 2 BACKGROUND

This study draws from practical background on GenAI, an emerging discourse on IT freelancers, SE4AI literature, and studies on the influence of technological hypes on development.

### 2.1 Current state of GenAI

In most general terms, GenAI embraces AI systems capable of generating new content, e.g., realistic images of people who never existed, based on examples [24]. Due to a stochastic layer, GenAI is produces nondeterministic output[1], so one cannot predict what image will emerge out of a specific set of examples which creates the illusion of creativity [5, 30]. The broad uptake of GenAI is driven by the availability of foundation models, i.e., general-purpose models trained on extensive data sets that cover a wide range of general tasks and can be adapted for more specific tasks [8]. Progress in natural language processing (NLP) yielded foundation models which accept natural text as input – LLMs. Following the November 2022 launch of OpenAI's ChatGPT, offering a chat interface for a powerful LLM, interest in GenAI skyrocketed. The initial hype around GPT models extended to other products relying on pretrained, externally hosted foundation models accessible through APIs [64]. Their rapid uptake captivated the scientific community, leading to numerous papers on potential implications for various sectors [18] and evaluating the models' suitability and quality for particular uses [7]. However, despite repeated calls, empirical data on GenAI's impact remains scarce.

An ecosystem of products has emerged around LLMs since late 2022. Major vendors, including OpenAI or Google offer access to hosted, closed-source models. Other providers offer open-source models, hosted on cloud infrastructure provided by HuggingFace or Replicate. Users can also train or fine-tune their own models though this requires expensive computational resources. LLMs can be accessed via dedicated APIs. Orchestration frameworks such as LangChain[2] or LlamaIndex[3] bundle these APIs and assist in pipelining functions. If an application depends on context data, these frameworks also support data preprocessing, storage, and management. A typical stack includes hardware or cloud resources, a LLM, its API, and optional orchestration framework and databases [30, 34]. Developers use this stack to create applications, typically using Python supported by major frameworks. Plugins for other languages and no-code platforms like Bubble.io make GenAI even more accessible. This ecosystem is highly dynamic.

This brief summary highlights that the development of GenAI-based applications extends far beyond simply prompting a model. More specifically, it underscores that GenAI, as a technology, is distinguished by its ability to generate novel, unseen content from examples or instructions in a non-deterministic fashion. It also refers to a variety of products, including models, APIs, and optional frameworks that provide access to these capabilities. Systems that leverage the unique capabilities of GenAI through these products are referred to as GenAI-based applications in this manuscript. The value of these applications is derived mostly from the customization of foundational models' capabilities for a specific task.

We still lack a clear understanding of the challenges that arise during the development of such applications, particularly when the GenAI components depend on pre-existing external models. Comprehensive reviews of the challenges associated with LLMs have only recently become available as preprints [e.g., 30, 34, 64]. They rely on analysis of grey and white literature showing that LLMs are subjects to the following limitations: need for large training data, tokenization problems, computational requirements, resources necessary for fine-tuning, inference latency, limited context length, model updating and refinement, bias, information hallucinations, lack of explainability, reasoning errors, risk of adversarial attacks or malign use, model behavior changes over time, as well as spelling/counting errors [30]. Further, results produced by LLMs cannot be reproduced even with decoding temperature set to zero [34]. While these technical issues and characteristics are widely confirmed, their effect on the work of independent developers remains unclear. Furthermore, the provided summaries overlook potential challenges that may arise in various subprocesses of SE when the LLMs are required to function as a component within a more complex system.

Natural-language prompting has become a central interaction mechanism with text-to-image and text-to-text foundation models. Prompts are instructions given to LLMs to make them follow instructions or standardize the output. They are the primary approach for users to control models' behavior. *Prompt engineering* has swiftly become a buzzword to describe the activity of crafting prompts for specific tasks [11, 63]. Scientific literature starts to depict activities involved in prompt engineering leading to nascent guidance or development of support tools [3, 54, 61–63]. Despite an emerging body of patterns and best practices, prompt engineering involves a significant amount of trial-and-error cycles with the success depending on the intuition of the prompter rather than systematic guidance [30, 63]. This is further amplified by prompt brittleness, i.e., unintuitive, large changes in the model output occurring due to very small changes in the prompt lake replacing words with synonyms or changing the order of requests [34]. Whereas those and other challenges mentioned above are documented in public discourse and in the most recent studies, we lack empirical data on how software engineers perceive those aspects and how the SE challenges of GenAI compare to the challenges known from non-generative machine learning (ML) technologies.

The impact of GenAI in software (SW) industry is currently a hot topic. Yet most conversations evolve around the pros and cons of using LLMs for code generation [13, 46, 61, 62]. Discussion about the use of them in applications remains a niche discourse limited, mostly, to specific application areas [52]. A proposed GenAI design process involves problem definition, model selection or model training, adaptation and alignment cycle (involving prompt engineering, fine-tuning, human alignment, and evaluation), and optimization and deployment [34, 64]. Although accurate, this advice is rather broad and lacks specific insights on how to tackle challenges unique to each phase. Moreover, to create value, GenAI technologies need to be integrated into either existing or new applications. This integration process encompasses standard SE steps, as well as design activities specific to GenAI. This could potentially complicate the development of GenAI-based applications, thereby introducing new SE challenges. This study aims to investigate these challenges from the viewpoint of freelance professionals.

---

[1] This holds for most GenAI including classic Generative Adversarial Networks (GANs) and LLMs, but some approaches such as Wasserstein GANs are deterministic.

[2] python.langchain.com, retrieved July 23, 2023
[3] gpt-index.readthedocs.io/en/latest/, retrieved July 23, 2023

## 2.2 Freelance developers

Freelancers, independent developers, and small agencies are actively engaging in the latest GenAI trends. A significant number of trending GitHub repositories related to GenAI are managed by individuals, not large tech corporations. Also, LLM orchestration frameworks such as LangChain and LlamaIndex are an outcome of independent and startup projects. Conversely, many client companies, lacking their own IT personnel and resources, are unable to delve into the potential of GenAI independently. As a result, they often turn to freelancers, viewing them as a cost-effective alternative to professional IT firms. This scenario places freelancers in a crucial role within the GenAI ecosystem.

Freelance developers, while integral to the IT industry, face numerous specific challenges such as inconsistent income, reliance on past clients' reviews, variable competition, evolving skill requirements, and dependency on freelancing platforms [28, 29]. They are often the first to be affected by shifts in trends and hypes [29], yet they play a pivotal role in the global delivery of IT development services and digital transformation [25, 26]. Existing research on freelancers in IT tends to take a broad approach, often overlooking the specific activities or project types these professionals engage in [26]. The subject of freelance developers is seldom addressed in SE research, and when it is, it is usually in the context of crowdsourced software development [1, 39]. Nevertheless, freelancers are increasingly serving as independent developers who can be directly contracted by client companies via freelancing platforms [28], handling more complex tasks than mere sporadic assistance [39]. We posit that the significance of freelancers will continue to grow with the advent of code generation models, enabling a single individual to efficiently develop an application or software module independently. However, this potential can only be fully realized if clients understand freelancers' viewpoints on emerging trends, thereby facilitating more effective collaboration. This has motivated our focus on freelance developers.

## 2.3 Development of AI-based applications

SE for AI (SE4AI) and SE for ML (SE4ML) are an expanding research domain that underscores the distinctive characteristics of AI-based development in contrast to conventional SW projects. Various studies [6, 15, 16, 31, 37, 38, 42, 57, 59] have indicated that uncertainty factors such as the probabilistic aspect of ML and dependency on big data present a multitude of issues for developers. A range of meta-studies summarizes the findings [23, 40, 42, 45]. Multiple studies attended to those challenges. Meanwhile, several comprehensive meta-reviews emerged listing and sorting the challenges according to ML process steps [47], SWEBOK knowledge areas [23, 42], or newly developed categories [44].

Many of the challenges listed in those studies overlap. Nevertheless, we also observed subtle differences in how they frame the challenges or how they order them. Given this observation, we combined four influential refereed meta-reviews published between 2021 and 2023 [23, 42, 44, 47] to obtain a good coverage of various challenges reported in the literature. We identified 137 distinct challenges provided in supplementary material [17][4]. Table 2 presents a selection of those challenges. To summarize, the main challenges stem from inadequate or absent data, necessity to engage in training as basic mode of creating AI capabilities, and the unpredictable and inherently complex nature of ML-based AI as evidenced by the *changing anything changes everything* (CACE) principle and models' opaqueness. The literature identifies lack of adequate guidance and process models to address those challenges, as well as problems with integrations of AI into end-user applications due to downstream compatibility problems caused by unpredictable or difficult to control output from AI components.

As explained above, GenAI contributes further technical challenges which the community just has started to understand and describe. This paper takes the perspective of freelance developers to understand the challenges perceive as particularly important and how they compare to SE4AI challenges. However, we speculate that technology is not the only source of challenges related to GenAI that freelancers need to deal with.

## 2.4 Technological hype and development

Despite the ongoing hype around GenAI [55] and other technologies, we found limited literature on SW development for hyped technologies and no recent comprehensive reviews. To provide background, we gathered literature from various fields, but the fragmented knowledge highlights the need for more research.

Hype occurs when a novel subject is heavily promoted across media platforms like newspapers, TV, and social networks to garner attention [10]. Since 1995, Gartner Inc., a consulting firm, has measured hype as fluctuating *expectations or visibility over time*, marking stages such as innovation trigger, peak of inflated expectations, trough of disillusionment, slope of enlightenment, and plateau of productivity [21, 53]. Typically, hype is used in relation to technologies reaching the peak of inflated expectations. Such technologies become fashionable, i.e., popular and socially demanded at this specific point in time [51]. Considering this discussion, we suggest defining technological hype as such: A technology experiences hype when it possesses inherent *novelty* (or includes novel features), thereby making it *fashionable* within business circles and society.

Hypes come and go. Various technologies gained attention in the past two decades. SE literature specifically attended to the implications of blockchain hype. The blockchain hype peaked in 2016-2017 [22]. With many platforms offering blockchain services, the focus of SE literature was on developing reliable smart contracts. However, traditional SW development life cycle models proved inadequate due to the immutability of smart contracts and the reliance on expert reviews over code testing [14, 43]. This triggered development of new techniques for Blockchain-Oriented SE to accommodate for the inherent nature of this technology [14, 49]. The hype around blockchain, rather than being solely negative, became an opportunity to problematize and improve SE practice and theory as pointed out by some researchers [12, 41]. Similarly, the hype around ML also garnered SE attention yielding insights listed above [12]. The concept of hype as a topic started entering SE discourse. Yet, it remains unclear how the fact that a technology is subject to hype impacts practitioners' work.

Agile development and its extensions, like MLOps or DevOps, were proposed to manage the novelty of technology and changing context. They became a standard across many companies [2]. Agile assumes that development can be divided into manageable parts, conflicting with the CACE principle. It also disregards the impact of hype, which significantly influences requirements. Despite these issues, agile remains influential in shaping SE practice.

---
[4] The supplementary material is available under https://doi.org/10.17605/osf.io/njc25

Hype around technology often increases demand for related solutions and prototypes. Driven by media coverage and success stories, companies seek to leverage these new technologies. This mirrors solution-based probing, where a technology is field-tested [9]. However, companies aim not just to learn, but also to create business value. The increased demand significantly affects the workforce, potentially leading to new developer profiles.

Two phenomena emerge due to this dynamic. The first, *resume-driven development*, occurs when firms and developers include trending technologies in job listings and resumes. This practice can foster harmful dynamics, compromise team reliability, and stress both employees and employers [19, 20]. The second, *hype-driven development*, often discussed by practitioners but not researchers [4, 33], occurs when developers use hyped technology without considering its architectural implications or integration effort. This particularly affects independent and agile developers.

Despite the recognition of hype's impact on SE practice, there is still limited research and guidance for freelancers navigating the technological opportunities and risks. Given the rapid pace of technological advancements, this topic warrants more attention. Future hypes are inevitable. Particularly, the specific challenges that arise when a technology, like GenAI, becomes a major hype, remain unclear. This study contributes to this by analyzing challenges related to GenAI and discussing the role of hype in them.

## 3 METHOD

This qualitative study identifies challenges reported by freelancers in GenAI projects. It uses a survey and interviews from 52 active freelance developers. The study was approved by the University's Institutional Review Board in May 2023. We collected data in May and June 2023 and analyzed it in June and July 2023. The study compares those issues against SE4AI / SE4ML challenges reported in earlier studies. We provide data concerning the method employed and the full list of challenges identified in SE4AI meta-reviews in supplementary material.

### 3.1 Literature Study

To compile a comprehensive, current list of challenges reported in SE4AI and SE4ML literature, we referred to four extensive meta-reviews [23, 42, 44, 47]. These were published in recognized, peer-reviewed outlets from 2021 to 2023, summarizing challenges related to AI-based application development reported by empirical studies. We extracted individual challenges from each meta-review, not only from tables or graphics but also from textual descriptions for a more detailed understanding. After removing duplicates, we identified 276 unique challenges with slight differences. We then grouped these challenges into 11 Knowledge Areas, including a new one, Training and Testing Data, based on the reviews [23, 42, 44]. Within these Knowledge Areas, we categorized the challenges into 54 groups, and after merging near-duplicates, we ended up with 137 challenges (see Table 2 and supplementary material [17]). These serve as a comparison basis with challenges mentioned by freelancers interviewed in this study.

### 3.2 Participants

We selected freelancers via Upwork, a freelancing platform. On May 1, we posted a job description offering $35 for a 10-minute survey and a 45-to-60-minute interview. We invited 450 freelancers (300 in May, 150 in June) who met specific criteria, including having a profile created before 2023, explicitly mentioning GenAI ('GPT', 'LLM', etc.) in their profile, and showing prior experience with the GenAI platforms and APIs. Following to the Upwork process, interested freelancers needed to submit a 'proposal' which, in our case, consisted of answers to four questions about their experience, GenAI knowledge, project history, and education. We received 96 answers. Some applicants had one or none GenAI projects while we required min. 2, provided nonsensical answers (e.g., empty ones), or proposed significantly higher rates, e.g., $1000. After filtering, we were left with 81 potential participants.

These individuals received a formal Upwork contract offer and survey invitation. However, 24 did not accept the offer or take the survey. Five participants withdrew from the study or became unresponsive after accepting the offer or filling out the survey. We excluded their responses from our analysis. Ultimately, 52 freelancers participated in the study.

Our participant group was diverse, as detailed in Table 1. The mean age was 32.6 years. On average, respondents had participated in 4.8 GenAI-related projects, a reasonable figure considering GenAI's broad uptake began in late 2022. Their average experience as developers was 6.9 years, be it as freelancer or in other settings. Most held a bachelor's or master's degree, with 36 specializing in informatics-related subjects. They reported using 14 different models/platforms, with 52 using GPT and 21 using DALL-E. They came from 23 countries across all inhabited continents. Unfortunately, gender diversity was low, with only two females despite our efforts to attract more.

**Table 1. Descriptive statistics regarding study participants**

| Attribute | Values and frequency among the participants |
| --- | --- |
| Age (years) | average: 32.6 years old, median: 31, max: 55, min: 19<br>*≤24 y.o.*: 11, *25-29*: 5, *30-34*: 17, *35-39*: 11, *40-49*: 5, *≥50*: 3 |
| Country of Residence | overall, 23 different countries including<br>*USA*: 7, *Pakistan*: 6, *India*: 5, *Nigeria*: 4, *Serbia*: 4, *Other*: 26 |
| No. GenAI projects | average: 4.8 projects on GenAI, median: 4, max: 15, min: 2<br>*2 projects*: 5, *3*: 11, *4*: 11, *5*: 13, *6*: 4, *7*: 4, *8*: 1, *≥10*: 3 |
| Developer experience | average: 6.9 years of experience, median: 5, max: 20, min: 1<br>*≤2 years*: 8, *3-4*: 10, *5-6*: 13, *7-8*: 5, *8-10*: 8, *≥10*: 8 |
| Degree | PhD: 6, MSc/MA: 21, BSc/BA: 21, Other: 1, BSc-Student: 3 |
| Subject studied | Overall, 20 distinct study subjects including<br>Computer Science: 18, ML / AI / NLP: 5, SE: 4, Electrical Engineering: 4, Physics: 3, Telecommunications: 2, Economics: 2, Geography: 2, IT Management/IS: 2. |
| GenAI platforms used | GPT: 52, DALL-E: 21, Stable Diffusion: 18, Midjourney: 17, LLaMA: 13, LaMDA/Bard: 6, Vicuna: 5, Others: 11; four participants mention fine-tuned or trained custom models |

### 3.3 Data Collection

We employed three data collection methods. Initially, participants provided basic demographic details with their proposal. After contract acceptance and study consent, they completed a survey on 2-5 chosen projects, and the main challenges and rewards of GenAI applications development. Then, they arranged an interview. The first author conducted these interviews, based on the problem-centered interview paradigm [58]. The data collected in the proposal and survey provided background for the exploration of the problems. The interview commenced with introduction and project background, followed by interviewees narrating their journey to becoming GenAI freelancers and discussing one or two selected projects that they were particularly proud of, or found challenging or intriguing. The elicited narratives were enhanced by dialogues featuring semi-structured questions related to freelancers' positive and negative experiences with GenAI, comparisons earlier and GenAI-based projects, project uncertainties, challenges, enjoyable aspects of working with GenAI, and their views on freelancing. The sequence of questions was adapted to the dialogue's flow. During the interview, the interviewer was referring back to challenges and projects participants described in the survey. The interviewer took notes to revisit crucial points from the narratives later in the interview.

The interviews were conducted via Zoom, averaging 55.7 minutes each (median: 55, min: 42, max: 81). All were in English and transcribed verbatim, then anonymized. The connection between demographic and other data was only possible through a token, ensuring privacy and security standards. We achieved data saturation – last 10 interviews did not introduce any significant new insights. We provide interview guides and other material related to data collection in the supplemental material.

### 3.4 Data Analysis

The data analysis process was iterative and combined thematic analysis during the initial coding followed interpretive analysis step during workshops [50, 56]. The second author conducted qualitative coding of the interview data in a bottom-up manner to avoid potential biases which could have emerged in the first author during interviews. He identified problems or issues, coding 268 segments as challenges. This process was supervised by all authors, with edge cases debated among them. After initial coding, two interpretation workshops were held. For a granular overview of challenges, they were distributed according to the SWEBOK Knowledge Areas [42] identified in the literature analysis of SE4AI meta-reviews (see Table 2 and [17]). We obtained multiple overlapping statements for all categories (except *Training and Testing Data*) confirming the saturation.

During interpretation workshops, the authors observed that freelancers differentiated between challenges that occurred because of the GenAI technology and because of the hype around it. For instance, freelancers predicted that some challenges might resolve at the end of the hype. Others were claimed to result from the attributes of GenAI as technology. After further exploration, authors were able to discriminate between challenges that were associated with the novelty and fashionableness of Gen AI as two dimensions of hype, and paradigm and product as two dimensions of technology. All authors agreed that this differentiation contributes to the understanding of collected insights. Supplementary material [17] provides further information about the coding schema and the freelancers quoted below.

### 4 RESULTS

Freelancers perceive multiple challenges distributed across 11 KAs and 54 categories as presented in Table 2. We count at least 99 distinct challenges related to development of applications based on GenAI of different granularity. Importantly, rather than identifying objective challenges, we focus on freelancers' perceptions. We want to understand what *they consider* unique, amplified, or reinforced due to the technological aspects of GenAI or the hype. This focus allows to produce guidance and identify development areas which address specific needs of this professional group. We first present the findings along the dimensions of technology and hype. In 4.5, we summarize the findings and present Table 2 which systematizes the findings.

### 4.1 GenAI Technology as Source of Challenges

**Characteristics of GenAI Paradigm**

Freelancers associate many challenges with the unique nature of GenAI. While some issues may be common to AI in general, GenAI's specific characteristics and accessibility create unique obstacles. LLMs are trained on vast datasets to produce high-quality, versatile outputs. However, these characteristics makes it impossible to trace and explain the model's behavior or identify the sources of a specific, problematic output. Freelancers indicate that this issue is exacerbated by many LLM providers' lack of transparency about training data sources and fine-tuning processes. For instance, it is unclear whether the providers used manual correction for specific use cases. This uncertainty affects both the content and form of the model's response. Freelancers say, they can control output's content by limiting the model to provided context data in retrieval augmented generation tasks which rely on processing of contextual information delivered to the model together with the prompt. Yet the response can still vary in tone, style, and structure. For instance, it might have inconsistent output format (bullet list vs. flow text). Measures to limit variability, such as GPT's 'temperature' parameter, are probabilistic, so freelancers report to observe different output for the same prompt even if temperature is set to minimum. The opaque relationship between a prompt and the output requires freelancers to invest significant time in trial-and-error process for crafting prompts:

> *I guess one of the most time-consuming things is to make it work. AI is a bit random and prompt engineering quite takes some time. And at that moment when you're like, 'okay, it works', some unstudied questions or something, you didn't try, [comes up] and you suddenly figure out that it doesn't produce results you wanted. (va-19-3)[5]*

A significant issue discussed by many freelancers are *hallucinations*. When prompted, GenAI models will attempt to generate an answer, even without relevant knowledge about the topic. This can result in incorrect or nonsensical text or images. Some freelancers feel uncomfortable delivering products without a guarantee of consistent, correct behavior. A seasoned freelancer explains that his dedication to high-quality products makes him the *slowest* part of the development process:

> *Well, the hallucinations, if the AI starts going off in creating information because one asked for it (...). Sometimes it's amusing, but other times it's like, 'Oh my God!' (...) It's lies. It's things that it's made up. There's no truthful basis to [check it] it, especially involving people. (...) So, I'm the slowest part in the process because I don't trust the AI. (...) So, I have to read everything and validate what it's doing. (...) Because, if you just take the output of the AI and trust it, it's going to bite you. (cl-55-15)*

Output hallucinations necessitate the establishment of guardrails, which may include filters or temperature reduction. However, this can limit GenAI's creativity which is desired for some use cases. A freelancer suggests that safety measures may conflict with client expectations, especially if the latter are unaware of GenAI's potential for hallucination:

---

[5] We use anonymous codes of the participants followed by their age and years of experience as developers to indicate the authors of the statements (xx-<age>-<experience>). For a detailed information about the participants, including the authors of those statements, see supplemental material (2.1.4 – 2.1.5).

*[There is] the idea of safety and kind of guardrails. (...) It's kind of a push-and-pull with certain clients. Because they say, 'oh, the text looks boring'. And, in my mind, I'm thinking, 'yeah, but it's better than the text looking crazy'. (...) Just because everything's looking good right now, doesn't mean things can't go off the edge. And, maybe, clients don't understand that, because, if you're doing a good job, they're never seeing it go off the edge. So, they don't know that the edge is there. (ev-38-13)*

Freelancers mention further challenges related to GenAI appearing during SW construction. Retrieval augmented generation applications rely on clients' data which provide context for LLMs to generate the answer. The intricate nature of LLMs, coupled with complex architecture and data, makes it exceedingly difficult for developers to identify the cause of incorrect or imprecise responses. A developer describes this issue as follows:

*For me, it's always about the data. (...) You get the data, you pump it in Pinecone and all that, and then start querying that. You realize at some point whatever the bot is giving you back is not as accurate as you wanted it to be. (...) Now you start asking yourself: is it the problem with the system, the workflow, the Pinecone, and querying setup problem? Is it a little problem? Or now you have to start thinking what could be the problem? So, for me, the data is the first step. (do-31-5)*

One of the consequences of these challenges is the emergence of downstream compatibility issues, which many freelancers find novel. The lack of control over the output, including its format, complicates software design. Freelancers must accommodate for variability through architectural decisions and focus on the interdependencies between components. This situation exacerbates reproducibility issues, rendering software testing less reliable and quality assurance more difficult. Furthermore, freelancers observe that large language models (LLMs) provide a set of interconnected advanced capabilities (e.g., evaluating and responding to a letter), making modularization unachievable. This complexity is a departure from earlier AI models, which were trained for individual capabilities. The opacity and scale of LLMs and other general AI technologies, combined with the absence of explanatory mechanisms, heighten the necessity for a trial-and-error approach.

Some interviewees consider training or fine-tuning as potential solution to some of the problems. Yet, they link the GenAI paradigm with enhanced expenses, high computational demands, and need for large volume of data beyond what was needed for earlier ML models. Only one interviewee mentioned having trained a model and claimed its cost higher than 10'000 USD for a single training round. For most freelancers such expenses are beyond the project budgets. This is a new aspect to those who worked with ML before and were used to train and re-train smaller models.

However, freelancers also indicate that available LLMs paradigm lessened the data and domain dependency. They speak of LLMs being able to deal with various data formats and structures for retrieval augmented generation. Also, they emphasize that progress they achieved in one project (e.g., prompt patterns) can be transferred easily to a project in a different domain, thanks to the generality of LLMs. This reduces pressure on SW construction and is ideal for freelancers who frequently move between projects.

**Characteristics of GenAI Products**

Many providers offer GenAI as a product or service, supplemented with APIs. This is the primary delivery method for LLMs. Despite offering easy access and a range of tools, freelancers report these models can also be unpredictably expensive. Providers adopt a pay-per-use strategy, where the cost per request is determined by the length of the prompt or response. This makes it challenging to estimate running costs for applications, particularly when user inputs are factored into the original prompt designed by the developer. This uncertainty also complicates budgeting for developers, as they struggle to gauge the economic resources required for the trial-and-error process and subsequent application testing. This complicates the SW configuration management and SE management. One freelancer elaborates on these budgeting challenges:

*You have a budget limit. So, you can't spend all the client's budget by doing a lot of prompts. You have to write it and to check it, but you don't have a lot of opportunities to rewrite, especially for big data. It's also a very difficult question and difficult issue because (...) if you if you have a coin and have 50 times tails, it doesn't mean that after the next one, you will get a tail. So, it's the same with GPT results because if you a dataset with 10 million [documents] and you will check manually 100 results, it doesn't mean everything was okay. (...) It's a kind of tradeoff because there are the limits, the time, and the money. And you can spend one of them, but you always have to check on the other. (ab-38-2)*

Further, freelancers identify unique technical constraints in LLMs which impact SW construction and maintenance. A commonly cited issue is the prompt or context window's size limits. There are strict restrictions on the number of tokens a single prompt can deliver (for example, 16,000 for GPT4), which might be insufficient. A freelancer who designed a search and summary engine for a large dataset explains this as follows:

*It is very important to find knowledge in the huge amount of data and you can take, for example, 1010 pages of a form with needed data and summarization for these ten pages. But for now, GPT can utilize only two pages and it is the limit. (eu-35-15)*

Software maintenance is hindered by the need to frequently adapt to model changes, a factor beyond the control of freelancers who depend on proprietary, closed-source models. Providers often update these models without transparency, leading to daily performance changes. This raises concerns about the consistency of model outputs, the predictability of application usefulness, and the need for ongoing re-validation of prompts and applications. This uncertainty also limits freelancers' project selection as they prefer to undertake projects, they can successfully deliver to garner positive client feedback and improve their completion rate. To mitigate this, freelancers engage in preselection activities like testing different prompts or a subset of client data before committing. However, their preselection decisions can be invalidated within days due to model changes, posing challenges due to increased dependency on external providers. Freelancers also worry about quality assurance, as they lack information on whether the external models are quality-assured, particularly in terms of security or privacy. The absence of this information from providers leaves freelancers' clients with doubts.

Despite obstacles, freelancers note that the emergence of robust foundational models has revolutionized their approach to delivering intelligent features. Previously, access to large, high-quality data sets, particularly in smaller projects, posed a significant hurdle in producing satisfactory products like free text chatbots. The introduction of LLMs has alleviated the strain associated with training and testing data, a key issue in previous AI generations.

### 4.2 GenAI Hype as Source of Challenges

**Novelty of GenAI**

The novelty of GenAI adds further challenges. Novelty amplifies and reinforces challenges related to unrealistic expectations from the clients and their limited knowledge. Lacking experience and reference projects, clients do not have a realistic grasp of GenAI's capabilities and do not know what developers can or cannot do.

This imposes extra effort on freelancers in the project's initial phase: instead of focusing on work, they must spend time elucidating and demonstrating GenAI's limitations which demands more communication with the clients and demands new explanation skills. A freelancer notes that clients sometimes question whether the problems depend on the technology or the freelancer.

> *In the end, this is a source of issues for me because they [clients] don't know what to expect. It seems like it [GPT] can do pretty much anything. [And if problems occur] they're still not sure: maybe it can do it, but I [the developer] am the one that cannot do it, or maybe it really is a technical limitation. It's very hard to communicate (...) what are the limitations. And moreover, why there are those limitations? (...) I try my best to explain it. And, so, I basically oversimplify. (ia-28-5)*

Many interviewees cite misinformation in public or social media as a problem. Clients often develop unrealistic expectations from such misleading information about what GenAI can achieve and how quickly. A developer, who is holds a PhD in computer science and has extensive experience with ML, explains:

> *I got a man who watched a video on YouTube, came to me, and said 'We need exactly this'. I said 'The thing you are referring is like a work of dozens of people for 3 to 6 months. And you want me to build this solution within seven days. So, this is not possible.' But he was insisting me to: 'If they can do this, then you must do too'. (ml-30-5)*

Another issue is the absence of a supportive community. While many freelancers appreciate the wealth of resources like courses, videos, and blogs, some struggle to locate the right community. As the GenAI becomes more popular and use cases solidify, this is expected to shift. This situation puts additional pressure on software engineering practices. The absence of a shared knowledge base and reference communities can hinder software development. A freelancer and start-up founder, who creates GenAI-based gaming applications, makes the following comparison:

> *For example, if you have any trouble with the Unity [a game engine platform], I'm sure you can point to the Unity Forum, and someone asked these questions. (...) I'm sure there are many little communities [around GenAI]. (...) There are a lot of people [engaged in them], but I'm not sure how many people worked in Unity integration with the generative AI and how can we find each other? I don't know. (tx-30-6)*

The unique nature of GenAI amplifies the challenge of unpredictable results, necessitating a trial-and-error method. With scarce experience, intuition, and lack of reference communities, freelancers struggle to assess a client's vision's feasibility. They experiment, boosting their initial, unpaid workload and stress. Given that clients choose freelancers based on early interactions, freelancers might end up working extended hours for no gain.

Freelancers concur that certain challenges may diminish over time. Initially, they anticipate accumulating knowledge that will simplify choosing appropriate clients and projects, confidently communicating with them, and establishing connections with reference communities. They believe that emerging metrics and quality assurance measures will validate their work's robustness. Secondly, they trust that potential clients will better understand their expectations and how to articulate their projects. Lastly, they predict that new tools, processes, orchestration frameworks, and additional resources will be introduced to aid in the development of applications and systems with GenAI components.

**Fashionableness of GenAI**

The fashionable character of GenAI developments present significant challenges. Older GenAI approaches like GANs, while available for around 10 years [24], have remained niche, used for specific cases. However, the arrival of GenAI relying on large, pre-trained models that accepts natural input has sparked new interest. Many companies are eager to utilize its seemingly unlimited capabilities hastily, often overlooking cheaper or simpler alternatives and not adequately examining their business needs. They are uncertain about their desired outcomes, making it challenging for them to define the metrics for evaluations. They also mistakenly assume that domain-dependent applications can be constructed without domain-specific data, expecting that LLMs possess access to all conceivable knowledge. An interviewee from a freelance agency emphasized that the hype exacerbates the problem of insufficient and unrealistic expectations, increasing pressure on processes associated with software requirements engineering and customer communication:

> *I think the biggest challenge from a non-technical perspective is understanding the use case. Many times, people would rush into [GenAI], 'Yeah, I want to chat about something. I want a generative AI solution.', but their problem can be solved with a much simpler solution. And, I think, it's really about understanding the problem before jumping into the solution instead of starting with the solution and trying to find the problem that fits for it. (ad-30-10)*

However, some clients limit access to end-users and other resources necessary for requirements engineering driven by the intention to obtain their solution in the shortest time possible. This has implications for subsequent phases of the SE process and might result in dissatisfied clients and, consequently, freelancers' low ranking. Therefore, some freelancers try to avoid potential clients who appear too much hype driven.

The surge in interest has additional implications. The rise in requests puts freelancers in a position of choice. This demand and accessibility of GenAI capabilities also entices less-experienced developers to enter the market. On the one hand, it increases the price competition as new freelancers try to attract clients through low prices. On the other hand, new freelance developers frequently lack the necessary skills to produce high-quality SW, but the high demand assures that they will be booked for projects. Some freelancers take on projects previously developed by others. An interviewee explains how he had to navigate these issues:

> *The work done on LangChain and OpenAI on Upwork is by developers that are not as experienced. Oftentimes clients end up with spaghetti code or some issue that they struggle with. (...) They got one file of Python, maybe a thousand lines of code, something like this. And I sat down, and I refactored it, and then I split into 6 or 7 files. (jf-23-5)*

This observation pertains to anti-patterns previously identified in standard SE and non-generative ML. Yet, the hype tends to draw freelancers who perpetuate these anti-patterns. These become problematic when there is a need to expand the application or assume responsibility for its maintenance and updates.

The ecosystem around generative models is incredibly complex and dynamic, with new solutions constantly emerging. This enhances pressure on freelancers, who often find their newly created solutions are rendered obsolete by new off-the-shelf tools, frameworks, or APIs. Moreover, they struggle to keep pace with the frequent technological changes. While upskilling is crucial for success, the unprecedented rate of change demands daily effort to stay updated and flexibility to adopt new solutions quickly. The dynamics underscores the importance of sensemaking and staying updated. Even seasoned freelancers perceive this rate of change as something distinct to the buzz around GenAI. An experienced developer illustrates how this influences his daily routine:

> *We're right on the edge of trying to keep up. And especially if you're looking at multiple solutions like I do, not just OpenAI, but Google and then, IBM, - it's a lot. There's so much coming out. And then, the cloud* *vendors and supporting vendors, they're releasing like crazy. And then, all of their AI-integrated products are hitting the market almost every day. (...). So, I'm still trying to catch up on their press release. And then boom, here comes AWS. You know, it's easily a couple of hours a day just trying to stay current with everybody. (...) In the early morning, I go through my LinkedIn connections, and then go through my emails, and my calendar. And then around noon I'll repeat that. And then just before dinner. And then, sometime between 10 and 11 at night. (cl-55-15)*

The rising GenAI trend increases pressure on freelancers, leading to issues in requirements engineering, professional practice, and SE management. Freelancers must skillfully manage their clients and invest significant effort to match technology with client needs. Even experienced freelancers find this hype novel and unprecedented as it magnifies older challenges and yields new ones.

### 4.3 Summary

Freelancers face many struggles when developing GenAI-based applications. Many appear new to them. Other seem aggravated compared to ML or their previous experience. However, freelancers also indicate that some old challenges get less important.

Based on the opinions collected from the freelancers and the literature covering SE4AI / SE4ML challenges, we compiled Table 2 to summarize our results and compare them against earlier studies. Literature refers to general challenges of developers while our results present freelancers' standpoints. Specifically, columns three and four in the table reflect statements from our data and our interpretation thereof. We use the last column in the table to indicate factors with primary influence on the challenges subsumed under a specific category. For instance, freelancers indicated fashionableness and novelty as major factors amplifying and aggravating challenges related to *customer expectations*, which we indicate with ↑. The star, ★, points to categories for which freelancers identified unique or new challenges. ↗ indicate which factors contribute to upholding challenges regarding GenAI in the category. For instance, novelty of GenAI results in clients' lack of knowledge about its capabilities. While freelancers mention this aspect as important, from their perspective similar problems were occurring earlier with ML. ↘ informs when a factor lessened the impact of challenges previously reported for AI/ML. This pertains especially to novel technological characteristics of LLMs which resolve problems of earlier ML-based solutions.

**Table 2. Comparison between challenges and needs reported in SE4AI literature and challenges and needs expressed by freelancers in our study along with indication of sources sorted by SWEBOK KAs and identified subcategories. See supplementary material [17] for the full version of this table with further distinction between singular challenges' impact factors.**

| KA | Subcategory | Challenges and Needs in SE4AI / SE4ML Literature [23, 42, 44, 47] | Challenges and Needs Mentioned by Freelancers (FL) | Impact Factors |
|---|---|---|---|---|
| Software Requirements | Customers' Expectations | • unrealistic 100% accuracy expectations, • unrealistic expectations toward adoption (e.g., running system with too little data) | • clients have too high expectations towards abilities of AI because of the hype around GenAI, • clients demand use of GenAI despite mismatch between the business requirements and GenAI's capabilities | ↑ fashion ↑ novelty |
| | Customers' Limited Knowledge | • lack of literacy concerning the capabilities of AI, • lack of knowledge on quantitative metrics | • clients use demanding/unrealistic projects to learn about limitations of GenAI generating risks for FL's rating, • new non-technical clients request GenAI-based solutions due to inflated expectations about GenAI's capabilities | ↗ fashion ↗ novelty |
| | Metrics vs. Requirements | • statistical metrics do not match requirements, • statistical metrics do not match business metrics, • difficult to use requirements coverage method, • lack of coverage-oriented datasets, • no operationalization of coverage for ML | • clients and FL lack effective statistical measures for quality assessment of generated content (as opposed, e.g., to precision and recall measures for classification tasks), • assessment of generated content requires domain expertise, • reliable ground truth is very difficult / impossible to create so evaluation cannot be reliably automated, • clients lack adequate business or quality criteria for new tasks if they haven't been previously conducted by humans | ★ paradigm ↑ fashion ↑ novelty |
| Software Design | Components Orchestration | • difficult dependencies between all parts of ML-based system, • additional complexity due to distributed architecture, • hard-to-manage interactions between ML models, | • orchestration was very difficult after release of some LLMs due to their novelty but lessened with new orchestration frameworks, • downstream compatibility is hard to achieve and maintain due to nondeterminism of GenAI's output format | ★ paradigm ↑ novelty |
| | Inherent Complexity of ML | • changing anything changes everything, • paradigm shift to pipeline-driven / system-wide view, • entanglement created by ML models, • abstraction boundary erosion | • singular LLM's capabilities (e.g., generating answer vs. generating summary) are impossible to separate technically because they rely on a single model, • similar or seemingly identical prompts can trigger different capabilities based on provision of different context data or subtle differences between prompts, • outputs of GenAI cannot be meaningfully explained or interpreted (black box) | ★ paradigm |
| | (Anti-)Patterns | • new anti-patterns due to complexity of ML including glue code, pipeline jungles, correction cascades, dead experimental code paths, technical debt, • lack of good patterns | • anti-patterns including glue code and correction cascades emerge due to nondeterminism of the LLMs' output's format, • good patterns are lacking due to new characteristics of LLMs, • hype attracts freelancers who reinforce anti-patterns due to lack of development experience | ↗ fashion ↗ paradigm ↗ product ↗ novelty |
| Software Construction | Data Dependency | • dependency on quality and availability of data during training, • continuous validation necessary, • poor calibration of data and models | • dependency on quality and availability of training data lessened due to availability of pre-trained models | ↘ product |
| | Domain Dependency | • reusing models across domains and contexts difficult, • transferring systems, • inability of models to generalize beyond training set | • problems related to transfer, generalizability, and reuse of models lessened due to generality and domain-independency of the available LLMs | ↘ paradigm ↘ product |
| | Insufficient Knowledge | • lack of understanding of models, libraries, and techniques, • lack of insight on problem at hand | • shared knowledge base and reference communities to obtain this knowledge are missing for some application scenarios due to novelty | ↑ novelty |
| | Integration | • presence of incompatible programming languages, • bugs in libraries, frameworks, and platforms, • interoperability problems | • bugs are present in database platforms (pinecone), orchestration frameworks, APIs due to their novelty, • frameworks or APIs undergo frequent, unannounced updates to turning developed pipelines ineffective, • context window and prompt length are limited in the | ↑ product |
| | | | currently available LLMs and APIs, • high latency times slow down the development and reduce user experience due to high load on externally hosted LLMs | |
| | Trial and Error | • problems in evaluating and debugging models, • training models requires many iterations, • analyzing and understanding structure and behavior of ML models, | • trial-and-error remains necessary but relates to analyzing existing models through prompting rather than training new models, • GenAI's reliance on billions of parameters makes trial-and-error for training and fine-tuning more costly and time-consuming, • need for trial-and-error increased because models more complex and untransparent concerning, e.g., data used for training | ↗ paradigm ↗ product |
| | Unreliable Output | • difficult control of quality during development of applications and models, • analyzing and understanding the structure and behavior of the model | • hallucinations occur in LLMs' output, • output's format and content are inconsistent, • output seems random and unpredictable, • output has low quality due to reasoning mistakes, • sources used by the model for answer generation are unclear and lack references, • debugging is more difficult due to unreliable output | ↑ paradigm |
| Software Testing | Experiment Complexity | • difficult to troubleshoot propagated data errors, • inherent variability behind parameters, • difficult to trace results due to long experiments and complex interactions | • inconsistency and variety of results from models make experiments more complex and require more experiments, • models' capabilities can change daily (e.g., models might 'unlearn' some capabilities) due to constant training / learning, • tracking experiment results is more difficult and not supported by APIs or frameworks | ↗ paradigm ↑ product |
| | Reproducibility | • hard to reproduce bugs and results due to non-determinism of results, • models' behaviors cannot be completely predicted making validation and verification difficult | • inherent nondeterminism of most GenAI paradigms like LLMs, which can be only partially reduced (e.g., temperature) but never turned off, makes test results hard to reproduce, • missing reproducibility enhances where user generated or dynamically collected data is used as context for a prompt due to potential interaction between context and prompt, • LLMs exhibit unstable performance with the same task | ↑ paradigm |
| | Testing Criteria | • lack of specification to test against, • discovery and definition of adversarial examples, • real-world testing difficult due to safety or security | • quantitative measures are lacking for generated content due to output's complex character, • content and form of output is less predictable than conventional ML due to generative nature, • opaqueness of models makes it difficult to specify adequate criteria for validation | ↑ paradigm |
| | Testing Data | • identification of test oracles, • missing access to high-quality test data, • manual labor necessary for creation of test data, • missing test cases | • manual creation of testing data is not feasible because many projects small and at the proof-of-concept stage, • testing data and test cases is missing due to customers' ad-hoc interest and innovative character of some projects | ↗ fashion |
| | Testing Tools and Support | • lack of test environments with trained ML models, • lack of cross-framework and cross-platform support, • lack of tools for testing | • specific tools for testing of GenAI technologies are lacking due to the novelty of the paradigm, • new testing processes are necessary for GenAI which are yet to come due to novelty of the paradigm | ↗ novelty |
| Software Quality | External Dependencies | • hard to understand the effects of external ML algorithms on desired qualities at runtime, • need to deal with not-assured components | • quality assurance of applications, which rely on external, non-assured models, is difficult due to lack of control over core functionality models, • quality assurance is difficult due to black box, untransparent nature of available LLMs | ↗ paradigm ↗ product |
| | Quality Standards and Criteria | • lack of criteria for the new types of quality features, • hard to define quality standards, • scalability of quality standards, • lack of certification, qualification, and standards on code quality | • effective and persistent guardrails for GenAI to assure the quality features are hard to specify due to variable output, • desired quality of output and possible limitations imposed by guardrails are hard to balance due to untransparent relation between input and output, • sufficient and adequate quality criteria for GenAI are missing (e.g., fairness metrics for LLMs' output) | ↗ paradigm |
| Software Maintenance | Inadequate Context of Use | • undeclared consumers who use system not in line with original purpose or intentions, • adaptation to changing environment of use, • inadequate user interfaces on top of ML systems | • difficult to predict or limit what input or context will be provided to the system by the user (if application allows for free user input, which most do) due to LLMs acceptability of unstructured input, • LLMs can handle unexpected input, but applications might inadequately process LLMs' output based on such input | ★ product |
| | Revalidation of Updated Models | • hard to determine frequency of training due to changes in context, • need to revalidate after updates, • updates in ML consider code, model, data as opposed to code in legacy systems, • changes in model behavior reduce users' trust and downstream/backward compatibility | • revalidation is hard due to frequent changes in the models delivered by external parties (e.g., OpenAI), • downstream compatibility is difficult to assure due to changes in the models and external dependencies, • applications' usability or usefulness difficult to guarantee over time as no information available on plans of the providers, • new requirements for validation since validation is necessary of the applications and prompts used therein and not the models themselves | ★ product |
| SW Configuration Mgmt | Economic and Computing Resources | • timing, memory, and energy constraints, • costs of training, • long training times for iterations, • need for adequate computing power | • new cost structure is necessary due to pay-per-use policy of major providers including costs of exploring external models' potentials and costs of using external models in the applications, • resources necessary for training of own GenAI models are incomparable to conventional ML due to models' complexity and size | ★ product |
| | Third-Party Dependency | *even though external dependencies are considered for quality assurance or development, considered literature does not explicitly attend to implications of external dependencies for software configuration management* | • dependency on external providers is significantly higher because applications rely on external models, access to APIs, and providers' infrastructure, • the usability and popularity of FL's applications depends on speed, availability, and compatibility of external components, • planning of economic resources is difficult due unclear plans and pricing strategies of external providers | ★ product |
| SE Process, Models, and Methods | Engineering | • need for continuous engineering, • necessity for ad-hoc development | • engineering is ad-hoc and results from short-term demand driven, e.g., by hype around GenAI | ↗ fashion |
| SE Professional Practice | Customer Communication | • hard to negotiate unfeasible expectations, • hard to convince clients to pay continuously for improvements, • need for skills to help customers set feasible targets / specify requirements | • common ground with clients from outside IT context is difficult due to hype and inflated expectations around GenAI, • communication with clients is difficult because clients rely on misinformation and inflated promises regarding GenAI | ↗ fashion |
| | End-user Engagement | • clients focus on technical solutions rather than involving end-users, • skepticism of users who were involved not early enough | • clients without experience in IT projects refuse to participate in requirements elicitation or provide requirements which hare not useful, • hype-driven clients expect fast results neglecting the need to involve end users in design | ↗ fashion |
| | Explanations | • difficult to explain to clients why models produce certain output, • insufficient explainability for some model types, • explaining the potential for improvement over time | • performance of LLMs and GenAI models difficult to explain to clients due to lack of knowledge about how models work, • opaqueness and inherent variability of the models make it difficult to understand output and make reliable statements, • clients blame freelancers for low performance of the models | ↑ paradigm |
| | Workforce Skills | • need for diverse skills, • lack of diverse skills in teams, • lack of adequately educated engineers | • skilled freelancers are lacking due to increased hype-driven demand for their skills and experience, • freelancers without adequate experience join the market due to enhanced demand and potential for lucrative jobs, • staying on top of dynamic market changes and technical developments causes major effort to keep pace with it | ↑ fashion |
| | Community and Competition | *the analyzed literature does not explicitly mention community-related topics as challenges* | • community to exchange with is difficult to find, especially on use of LLMs and GenAI for specific domains, due to novelty and niche character of some projects, • competition increases due to many new freelancers entering the market | ★ fashion ★ novelty |
| SE Management | Planning Uncertainty | • uncertainty of estimating development time and cost, • no well-defined input-output factor for validation, • assessment of long-term potential based on short-term metrics, • impossibility to make guarantees on cost-effectiveness | • estimations are difficult to make due to lack of experience, • limited ability to say upfront if a specific case can be covered by GenAI technologies due to lack of experience, • clients expect that a certain task will be solved with GenAI despite more adequate and easier-to-plan alternatives, e.g., conventional ML, • reasonable offer / putting a price tag on a project difficult to due to lack of heuristics and experience | ↑ novelty ↗ fashion ↗ paradigm |
| | Resources Shortage and Costs | • resource limits curbing down efforts, • lack of organizational incentives, • lack of resources | • costs related to experimentation during prompt-engineering are hard to manage and control especially if the client does not cover those costs | ↗ product |
| | Pre-Selection Effort | *the analyzed literature does not explicitly mention challenges related to pre-selection of projects* | • preselecting projects is hard based on the requests from clients due to lack of experience, • fashionableness causes many unusual or incomplete requests and assessing them requires effort to explore whether GenAI technologies are useful to solve the requests, • frequent changes and instability of the models might invalidate pre-selection decisions | ★ fashion ★ product ★ novelty |
| Training and Testing Data | Data Access | • hard to discover what data exists and where, • difficulty to collect new data and prepare it | • no need for training data and training data quality assurance due to the availability of pre-trained models *FL refer to challenges related to data if their projects involve (a) training of fine-tuning models or (b) retrieval augmented generation; in both cases the challenges confirm many of the challenges listed for conventional ML, larger datasets are necessary for training a generative model.* | ↘ product ↗ paradigm |
| | Data Management | • hard to control, version, and deploy data | | |
| | Data Preparation | • need to combine, transfer, clean, and transform data, • lack of tools for data preparation, • inconsistent data types and quality | | |
| | Data Quality | • difficult to assure data quality, • missing values and rare cases in data, • lack of tools to annotate and assure data quality, • bias in data | | |

## 5 DISCUSSION

Freelancers report on a multitude of challenges they experience when developing applications including GenAI components. They are surprised by the magnitude of some challenges, e.g., issues related to stay on top of changes, while others appear completely new to them, e.g. impracticality of separating functionalities. Given the increasing role of freelance developers in digital transformation [25–29, 39], their insights are invaluable to the broader community. Simultaneously, the results provide information about freelancers and their work, such as the importance of the adequate projects' selection. In the following, we interpret our findings and suggest avenues of research for SE4GenAI and for HypeSE.

### 5.1 Software Engineering for Generative AI

What does GenAI imply for freelancers? Freelancers suggest that GenAI equips them to develop potent applications within a constrained timeframe. Rather than engaging in expensive and lengthy processes of data collection, preprocessing, and training, they delve into prompts that can be efficiently handled by an individual. Freelancers from non-technical backgrounds express that GenAI has unlocked new avenues for them, fostering opportunities for economic and professional growth. They leverage freelance jobs to enhance their skills and render their freelance career more sustainable. However, they require practical guidance, illustrative examples, and consistent interaction to achieve this.

The freelancers would have appreciated more structured guidance on GenAI technology, both as a concept and a product. They reported unexpected issues stemming from hallucinations, challenges in handling context data during prompting, and complications associated with prompt engineering. Their pursuit of solutions and understanding of these phenomena across the Internet was often messy, leading to a disorganized process and outcome. Our own research indicates that comprehensive, professional guidance on potential challenges is in its infancy and fragmented across platforms and communities. [30, 34–36, 60, 64]. Freelancers' ability to improvise becomes a core asset in this situation. Short trial-and-error typical for working with GenAI turns *improvisation* into the actual value offering of freelancers. This, however, comes not without challenges for SE.

Freelancers are particularly affected by challenges related to characteristics of *GenAI paradigm* during *SW design*, *construction*, *testing*, and *quality* assurance due to hallucinations, low reproducibility, reasoning errors, inherently probabilistic output, and the CACE principle. Also, insufficient explainability leads to issues in contact with customers affecting freelancers' professional practice. Whereas the technical challenges are known [30, 34], our results documents their impact on day-to-day SE practice of freelancers.

The *characteristics of GenAI products* influence *SW construction*, *SW maintenance*, *SW configuration management*, and *SE management*. Factors influencing this situation include the constant changes and opaque update cycles of leading providers like OpenAI, the availability and latency of models (including not just inference latency, but also latencies from workload distribution), cost structures and policies enforced by providers, and the minimal control over model behavior. Currently, freelancers do not view any significant alternatives to these major providers due to the restricted capabilities of open-source models or the prohibitive costs of fine-tuning and training. This indicates that the question of whether GPT is all one needs [64] will remain valid for the years to come. The dependency on external providers and their products was a marginal issue in earlier SE4AI literature [23, 42]. Yet, the fact that training and hosting own GenAI models is frequently impracticable, the issue of managing external dependencies rises.

Based on freelancers' reports, we propose a new area of research, Software Engineering for Generative AI (SE4GenAI). We propose the following key components that SE4GenAI must incorporate: (1) systematic guidance for activities like managing the trial-and-error loop as the primary mode in making progress during the SW construction; (2) strategies and metrics for reliable testing, quality assurance, and downstream compatibility to handle GenAI's variable outputs and dynamic capability changes, including ways to make adequate guarantees concerning functionality; (3) guidance on dealing with external dependencies impacting functional and non-functional aspects of the system during maintenance as well as the cost structure of the product; (4) support for more effective explanation practices in communication with the clients about (at least currently) unexplainable and unpredictable model behavior. Equipped with that knowledge, freelancers in our study would have had a better chance to complete their products more efficiently.

## 5.2 Hype-Induced Software Engineering

What does *hype* mean for freelancers? The escalating excitement around GenAI has significantly influenced the dynamics of the freelance market. Our data clearly indicates that the success of freelancers is heavily dependent on job platforms and their respective communities. To maintain their success, freelancers must master the art of navigating these platforms, which may sometimes involve declining projects of personal interest if the likelihood of success seems uncertain. In this era of heightened interest, the key strengths of freelancers are their autonomy, agility, and flexibility. Particularly for those who work on short, time-bound projects, they can swiftly adapt to new technologies and capitalize on the current buzz. However, this also means that freelancers must be able to acquire new skills quickly and independently, a process that demands both time and effort.

The integration of GenAI and the escalating demand for developers has presented new challenges that have affected freelancers, their work routines, and even their personal lives. We hypothesize that the rapid advancement of technology will likely generate similar trends in the future, and the impact of these trends on the IT development sector will extend beyond freelancing. This could potentially disrupt many established SE practices and guidelines. We argue that the strategies employed by experienced freelancers to navigate the hype provides valuable insights for SE research. Concurrently, formalizing this knowledge about navigating trends could be beneficial for new freelancers seeking guidance.

*Fashionableness of GenAI* is primarily reflected in freelancers' comments concerning new or amplified challenges on *SW requirements*, *SE professional practice*, and *SE management*. The hype incites unexperienced clients and unexperienced, low-skilled freelancers to enter the market. Clients come with exacerbated, misinformed expectations and unclear or hard-to-measure targets. Frequently, they are simply interested in solution-based probing [9] to discover new opportunities. Unexperienced freelancers, even if trained as ML specialists or software developers, lack tools and skills to interact with such clients in a professional way.

Freelancers are now operating in a highly competitive environment that necessitates careful project selection and intense pre-contract solution testing. This challenge, which has not been addressed in previous SE4AI research [42] or SE hype literature [19], is substantial. We hypothesize that this will lead to a reciprocal effect on the client side, particularly among inexperienced businesses and startups. These entities must undertake a complex search for suitable talent, prompting freelancers to engage in resume-driven development to establish credibility [19, 20]. The dynamics of GenAI hype differ from previous trends such as blockchain. For instance, the development of smart contracts necessitated extensive client-engaged requirements engineering [14, 49], which made the freelancers' work transparent and accountable, compelling the client to contemplate their objectives. Moreover, blockchain's use cases were primarily confined to finance and value capture [12, 14, 41, 43, 49], whereas GenAI encompasses a wider range of applications. This compels freelancers to rapidly acquire the requisite domain expertise.

*Novelty of GenAI* further amplifies effects created by the GenAI's fashionable character concerning *SW requirements*, *SE management*, and *SE professional practice*. It increases planning uncertainty due to lack of heuristics, reference projects, and reference communities, which makes the overall situation harder. The novelty encourages ad-hoc efforts and quick fixes, anti-patterns in *SW design*, and inefficient communication in *SE practice*.

Considering the likelihood of future hypes, we advocate for research in Hype-Induced Software Engineering (HypeSE). Freelance and independent developers, due to their flexibility and agility, are most impacted by hypes and require support. HypeSE should incorporate: (1) guidance on managing inflated expectations and communicating these to clients and users, extending existing literature on requirements elicitation and management; (2) processes and tools for prioritizing and managing project opportunities to support economic stability and mitigate risks associated with riding high waves; (3) support to help developers manage incoming hype information, including fostering effective community building and facilitating expert information exchange; (4) definition of new roles such as domain-specialist freelancers who provide in-depth insight and consultancy to clients based on expertise collected in similar projects assuring diffusion of knowledge between client companies and enabling them to spot opportunities for clients. To accommodate for such new roles, we claim CS and SE education should focus on equipping future developers with specific domain skills rather than producing domain-agnostic generalists. This could foster directions like SE for the financial industry, life sciences, or social media.

### 5.3 Generalizability and Threats to Validity

While this study highlights the need for SE4GenAI and HypeSE based on freelancer experiences, we propose that other stakeholder groups and professions may face similar, related, or complementary challenges. Further research is required to comprehend these challenges, emphasizing the unique aspects of each stakeholder group. For example, issues related to imprecise or exaggerated software requirements may be interpreted differently by clients, contractors, and requirements engineers. Addressing these problems necessitates understanding the mechanisms that produce these challenges for each group. This viewpoint would supplement the currently prevalent approach, which attempts to aggregate software engineering challenges and their solutions at the technology level, such as GenAI [18, 30, 34, 63, 64] or AI/ML [23, 40, 42, 44, 47]. We recognize existing efforts to concentrate on developers as a stakeholder group [59] and specific studies focusing on particular areas of SE, including requirements engineering or testing [32, 48]. We advocate for the amplification of these efforts by incorporating the perspectives and specific contexts of project managers, client companies, ML model providers, and so on.

Interview studies risk low interpretation validity due to suggestive questioning, personal bias, and coder assumptions. We mitigated this by allowing interviewees to begin with an open narrative, capturing extensive information without a specific focus. An uninvolved researcher coded the data to minimize bias. We maintained a theory-agnostic approach to avoid guiding interviewees or analysis. However, we acknowledge the first author's prior involvement in AI-based studies may have influenced him.

This study's internal validity may be influenced by context and respondent biases, as it relies on reports potentially affected by recent events or second-hand reflections. The external validity could be impaired due to the nascent nature of GenAI, which is yet to stabilize. We mitigate this by distinguishing challenges linked to GenAI's characteristics from those with different origins. Additionally, we sought to improve triangulation by interviewing a diverse group of developers.

## 6 CONCLUSION AND FUTURE WORK

This study was conducted with the objective of pinpointing the challenges freelancers face in relation to GenAI. Freelancers, who serve as ideal collaborators for small to mid-sized enterprises exploring new technologies, provided insights into the challenges they perceive as new or significantly heightened due to the hype surrounding GenAI, as well as GenAI's inherent and product-related features. Freelancers have observed a shift in workload from data management and model training to model interaction and output management, which complicates their ability to rely on previous intuition and experience. Moreover, the hype around GenAI contributes to problems arising from overblown expectations and swift changes that demand the freelancer's attention. This situation parallels notable historical events such as the gold rush or the exploration of the wild west, where independent pioneers ventured into uncharted territories.

We believe the SE community should monitor future developments in this field. The findings of this study could be reinforced through additional qualitative research or large-scale confirmatory surveys. Ethnographic studies based on in-situ observation of larger development teams could offer insights into how freelancers tackle practical challenges and uncover unexplored issues. We also recommend normative and prescriptive research to develop new standards and techniques for software engineers and managers.

The study provides practitioners with insights into the challenges of SE for GenAI-based solutions. The provided list of SE challenges is thought to sensitize them for specific issues to emerge. The study also educates practitioners about the impact and risks of hype. Especially freelancers depending on external contracts are subject to those risks. SE researchers gain an empirical basis for deriving guidance for practitioners and become more aware of hype as an inherent element in SE. Its influence extends beyond the context of GenAI and directly affects SE practice. We suggest addressing these challenges through intensified research on SE4GenAI and HypeSE.

### ACKNOWLEDGEMENTS

We would like to express our gratitude to all the participants of the study for their valuable time, dedicated efforts, and keen interest. This work has been funded by the authors' University itself.

### REFERENCES

[1]  Abhinav, K. and Dubey, A. 2017. Predicting budget for Crowdsourced and Freelance Software development Projects. *Proc. Innovations in Software Engineering Conf.* (New York, NY, USA, Feb. 2017), 165–171.
[2]  Agile Adoption Report 2022: 2022. *https://certiprof.com/pages/certiprof-agile-adoption-report-2022.* Accessed: 2023-07-23.
[3]  Arawjo, I., Vaithilingam, P., Wattenberg, M. and Glassman, E. 2023. ChainForge: An open-source visual programming environment for prompt engineering. *Adjunct Proc. Annual ACM Symposium on User Interface Software and Technology* (New York, NY, USA, Oktober 2023), 1–3.
[4]  Are You a Hype-Driven Developer? 2020. *https://betterprogramming.pub/are-you-a-hype-driven-developer-368586c27051.* Accessed: 2023-07-23.
[5]  Arjovsky, M., Chintala, S. and Bottou, L. 2017. Wasserstein GAN. arXiv. http://arxiv.org/abs/1701.07875
[6]  Arpteg, A., Brinne, B., Crnkovic-Friis, L. and Bosch, J. 2018. Software Engineering Challenges of Deep Learning. *Proc. Euromicro Conf. on Software Engineering and Advanced Applications* (Prague, Czech Republic, Aug. 2018), 50–59.
[7]  Ayers, J.W. et al. 2023. Comparing Physician and Artificial Intelligence Chatbot Responses to Patient Questions Posted to a Public Social Media Forum. *JAMA Internal Medicine.* 183, 6 (Jun. 2023), 589–596.
[8]  Bommasani, R. et al. 2022. On the Opportunities and Risks of Foundation Models. arXiv. http://arxiv.org/abs/2108.07258
[9]  Briggs, R.O., Böhmann, T., Schwabe, G. and Tuunanen, T. 2019. Advancing Design Science Research with Solution-based Probing. *Proc. Hawaii International Conference on System Sciences* (Grans Wailea, Hawaii, Jan. 2019).
[10] Cambridge Dictionary 2023. hype. *Cambridge Dictionary.* Cambridge UP.
[11] Chen, B., Zhang, Z., Langrené, N. and Zhu, S. 2023. Unleashing the potential of prompt engineering in Large Language Models: a comprehensive review. arXiv. http://arxiv.org/abs/2310.14735

[12] Colomo-Palacios, R. 2020. Cross Fertilization in Software Engineering. *Systems, Software and Services Process Improvement* (Cham, 2020), 3–13.
[13] Destefanis, G., Bartolucci, S. and Ortu, M. 2023. A Preliminary Analysis on the Code Generation Capabilities of GPT-3.5 and Bard AI Models for Java Functions. arXiv. http://arxiv.org/abs/2305.09402
[14] Destefanis, G., Marchesi, M., Ortu, M., Tonelli, R., Bracciali, A. and Hierons, R. 2018. Smart contracts vulnerabilities: a call for blockchain software engineering? *Proc. Intl. Workshop on Blockchain Oriented Software Engineering* (Campobasso, Italy, Mar. 2018), 19–25.
[15] Dolata, M. and Crowston, K. 2024. Making sense of AI systems development. *IEEE Transactions on Software Engineering*. 50, 1 (2024).
[16] Dolata, M., Crowston, K. and Schwabe, G. 2022. Project Archetypes: A Blessing and a Curse for AI Development. *Proc. Intl. Conf. on Information Systems* (Dec. 2022).
[17] Dolata, M., Lange, N. and Schwabe, G. 2023. Supplementary Material: Dolata et al. "Development in times of hype." OSF. https://doi.org/10.17605/OSF.IO/njc25
[18] Dwivedi, Y.K. et al. 2023. Opinion Paper: "So what if ChatGPT wrote it?" Multidisciplinary perspectives on opportunities, challenges and implications of generative conversational AI for research, practice and policy. *International Journal of Information Management*. 71, (Aug. 2023), 102642.
[19] Fritzsch, J., Wyrich, M., Bogner, J. and Wagner, S. 2023. Resist the Hype! Practical Recommendations to Cope With Résumé-Driven Development. *IEEE Software*. (2023), 1–8.
[20] Fritzsch, J., Wyrich, M., Bogner, J. and Wagner, S. 2021. Résumé-Driven Development: A Definition and Empirical Characterization. *Proc. Intl. Conf. on Software Engineering: Software Engineering in Society* (May 2021), 19–28.
[21] Gartner Hype Cycle Research Methodology: 2023. *https://www.gartner.com/en/research/methodologies/gartner-hype-cycle*. Accessed: 2023-07-23.
[22] Gayvoronskaya, T. and Meinel, C. 2021. *Blockchain: Hype or Innovation*. Springer International Publishing.
[23] Giray, G. 2021. A software engineering perspective on engineering machine learning systems: State of the art and challenges. *Journal of Systems and Software*. 180, (Oct. 2021), 111031.
[24] Goodfellow, I.J., Pouget-Abadie, J., Mirza, M., Xu, B., Warde-Farley, D., Ozair, S., Courville, A. and Bengio, Y. 2014. Generative Adversarial Networks. arXiv. http://arxiv.org/abs/1406.2661
[25] Gupta, V., Fernandez-Crehuet, J.M., Gupta, C. and Hanne, T. 2020. Freelancing Models for Fostering Innovation and Problem Solving in Software Startups: An Empirical Comparative Study. *Sustainability*. 12, 23 (Jan. 2020), 10106.
[26] Gupta, V., Fernandez-Crehuet, J.M. and Hanne, T. 2020. Freelancers in the Software Development Process: A Systematic Mapping Study. *Processes*. 8, 10 (Oct. 2020), 1215.
[27] Gupta, V., Fernandez-Crehuet, J.M., Hanne, T. and Telesko, R. 2020. Fostering product innovations in software startups through freelancer supported requirement engineering. *Results in Engineering*. 8, (Dec. 2020), 100175.
[28] Gussek, L., Grabbe, A. and Wiesche, M. 2023. Challenges of IT freelancers on digital labor platforms: A topic model approach. *Electronic Markets*. 33, 1 (Oct. 2023), 55.
[29] Gussek, L. and Wiesche, M. 2023. IT Professionals in the Gig Economy. *Business & Information Systems Engineering*. 65, 5 (Oct. 2023), 555–575.
[30] Hadi, M.U., Tashi, Q.A., Qureshi, R., Shah, A., Muneer, A., Irfan, M., Zafar, A., Shaikh, M.B., Akhtar, N., Wu, J. and Mirjalili, S. 2023. Large Language Models: A Comprehensive Survey of its Applications, Challenges, Limitations, and Future Prospects. TechRxiv. https://www.techrxiv.org/users/618307/articles/682263
[31] Hesenius, M., Schwenzfeier, N., Meyer, O., Koop, W. and Gruhn, V. 2019. Towards a Software Engineering Process for Developing Data-Driven Applications. *Proc. Intl. Workshop on Realizing Artificial Intelligence Synergies in Software Engineering* (Montreal, QC, Canada, May 2019), 35–41.
[32] Huang, S., Liu, E.-H., Hui, Z.-W., Tang, S.-Q. and Zhang, and S.-J. 2018. Challenges of Testing Machine Learning Applications. *International Journal of Performability Engineering*. 14, 6 (Jun. 2018), 1275.
[33] Hype Driven Development: 2017. *https://blog.daftcode.pl/hype-driven-development-3469fc2e9b22*. Accessed: 2023-07-23.
[34] Kaddour, J., Harris, J., Mozes, M., Bradley, H., Raileanu, R. and McHardy, R. 2023. Challenges and Applications of Large Language Models. arXiv. http://arxiv.org/abs/2307.10169
[35] Katsiuba, D., Kew, T., Dolata, M., Gurica, M. and Schwabe, G. 2023. Artificially Human: Examining the Potential of Text-generating Technologies in Online Customer Feedback Management. *Proc. Intl. Conf. Information Systems* (2023).
[36] Katsiuba, D., Kew, T., Dolata, M. and Schwabe, G. 2022. Supporting Online Customer Feedback Management with Automatic Review Response Generation. *Proc. Hawaii International Conference on System Sciences* (2022).
[37] Khomh, F., Adams, B., Cheng, J., Fokaefs, M. and Antoniol, G. 2018. Software Engineering for Machine-Learning Applications: The Road Ahead. *IEEE Software*. 35, 5 (Sep. 2018), 81–84.
[38] Kim, M., Zimmermann, T., DeLine, R. and Begel, A. 2018. Data Scientists in Software Teams: State of the Art and Challenges. *IEEE Transactions on Software Engineering*. 44, 11 (Nov. 2018), 1024–1038.
[39] LaToza, T.D. and van der Hoek, A. 2016. Crowdsourcing in Software Engineering: Models, Motivations, and Challenges. *IEEE Software*. 33, 1 (Jan. 2016), 74–80.
[40] Lorenzoni, G., Alencar, P., Nascimento, N. and Cowan, D. 2021. Machine Learning Model Development from a Software Engineering Perspective: A Systematic Literature Review. arXiv. http://arxiv.org/abs/2102.07574
[41] Marchesi, M. 2018. Why blockchain is important for software developers, and why software engineering is important for blockchain software (Keynote). *Proc. Intl. Workshop on Blockchain Oriented Software Engineering* (Campobasso, Italy, Mar. 2018), 1–1.
[42] Martínez-Fernández, S., Bogner, J., Franch, X., Oriol, M., Siebert, J., Trendowicz, A., Vollmer, A.M. and Wagner, S. 2022. Software Engineering for AI-Based Systems: A Survey. *ACM Transactions on Software Engineering and Methodology*. 31, 2 (Apr. 2022), 37e:1-37e:59.
[43] Miraz, M.H. and Ali, M. 2020. Blockchain Enabled Smart Contract Based Applications: Deficiencies with the Software Development Life Cycle Models. arXiv. http://arxiv.org/abs/2001.10589
[44] Nahar, N., Zhang, H., Lewis, G., Zhou, S. and Kästner, C. 2023. A Meta-Summary of Challenges in Building Products with ML Components -- Collecting Experiences from 4758+ Practitioners. *Proc. Intl. Conf. on AI Engineering - Software Engineering for AI* (Melbourne, Australia, Mar. 2023).
[45] Nascimento, E., Nguyen-Duc, A., Sundbø, I. and Conte, T. 2020. Software engineering for artificial intelligence and machine learning software: A systematic literature review. arXiv. http://arxiv.org/abs/2011.03751
[46] Nguyen-Duc, A. et al. 2023. Generative Artificial Intelligence for Software Engineering -- A Research Agenda. arXiv. http://arxiv.org/abs/2310.18648
[47] Paleyes, A., Urma, R.-G. and Lawrence, N.D. 2022. Challenges in Deploying Machine Learning: A Survey of Case Studies. *ACM Computing Surveys*. 55, 6 (Dezember 2022), 114:1-114:29.
[48] Pei, Z., Liu, L., Wang, C. and Wang, J. 2022. Requirements Engineering for Machine Learning: A Review and Reflection. *Proc. Intl. Requirements Engineering Conf. Workshops* (Melbourne, Australia, Aug. 2022), 166–175.
[49] Porru, S., Pinna, A., Marchesi, M. and Tonelli, R. 2017. Blockchain-Oriented Software Engineering: Challenges and New Directions. *Proc. Intl. Conf. on Software Engineering: Companion* (Buenos Aires, Argentina, May 2017), 169–171.
[50] Schwartz-Shea, P. and Yanow, D. 2012. *Interpretive research design: concepts and processes*. Routledge.
[51] Simmel, G. 2020. Fashion. *Fashion Theory*. Routledge.
[52] Stappen, L., Dillmann, J., Striegel, S., Vögel, H.-J., Flores-Herr, N. and Schuller, B.W. 2023. Integrating Generative Artificial Intelligence in Intelligent Vehicle Systems. arXiv. http://arxiv.org/abs/2305.17137
[53] Steinert, M. and Leifer, L. 2010. Scrutinizing Gartner's hype cycle approach. *Proc. Portland Intl. Conf. on Management of Engineering and Technology* (Portland, Oregon, USA, Jul. 2010), 1–13.
[54] Strobelt, H., Webson, A., Sanh, V., Hoover, B., Beyer, J., Pfister, H. and Rush, A.M. 2023. Interactive and Visual Prompt Engineering for Ad-hoc Task Adaptation with Large Language Models. *IEEE Transactions on Visualization and Computer Graphics*. 29, 1 (Jan. 2023), 1146–1156.
[55] Vinsel, L. 2023. Don't Get Distracted by the Hype Around Generative AI. *MIT Sloan Management Review*. 64, 3 (2023), 1–3.
[56] Walsham, G. 1995. Interpretive case studies in IS research: nature and method. *European Journal of Information Systems*. 4, 2 (May 1995), 74–81.
[57] Wan, Z., Xia, X., Lo, D. and Murphy, G.C. 2021. How does Machine Learning Change Software Development Practices? *IEEE Transactions on Software Engineering*. 47, 9 (Sep. 2021), 1857–1871.
[58] Witzel, A. and Reiter, H. 2012. *The Problem-Centred Interview: Principles and Practice*. SAGE Publications Ltd.
[59] Wolf, C.T. and Paine, D. 2020. Sensemaking Practices in the Everyday Work of AI/ML Software Engineering. *Proc. Intl. Conf. on Software Engineering: Workshops* (New York, NY, USA, Sep. 2020), 86–92.
[60] Wu, W., Heierli, J., Meisterhans, M., Moser, A., Färber, A., Dolata, M., Gavagnin, E., de Spindler, A. and Schwabe, G. 2023. PROMISE: A Framework for Model-Driven Stateful Prompt Orchestration. arXiv. http://arxiv.org/abs/2312.03699
[61] Xu, J., Cui, Z., Zhao, Y., Zhang, X., He, S., He, P., Li, L., Kang, Y., Lin, Q., Dang, Y., Rajmohan, S. and Zhang, D. 2024. UniLog: Automatic Logging via LLM and In-Context Learning. *Proc. Intl. Conf. Software Engineering* (New York, NY, US, Apr. 2024), 129–140.
[62] Yu, H., Shen, B., Ran, D., Zhang, J., Zhang, Q., Ma, Y., Liang, G., Li, Y., Wang, Q. and Xie, T. 2024. CoderEval: A Benchmark of Pragmatic Code Generation with Generative Pre-trained Models. *Proc. Intl. Conf. Software Engineering* (New York, NY, US, Apr. 2024), 417–428.
[63] Zamfirescu-Pereira, J.D., Wong, R.Y., Hartmann, B. and Yang, Q. 2023. Why Johnny Can't Prompt: How Non-AI Experts Try (and Fail) to Design LLM Prompts. *Proc. Conf. on Human Factors in Computing Systems* (New York, NY, USA, Apr. 2023), 1–21.
[64] Zhang, C. et al. 2023. A Complete Survey on Generative AI (AIGC): Is ChatGPT from GPT-4 to GPT-5 All You Need? arXiv. http://arxiv.org/abs/2303.11717

# Supplementary Material

## Overview

This is the supplementary material for the manuscript "Development in times of hype: How freelancers explore Generative AI?" by Dolata, Lange, and Schwabe accepted for publication at ICSE 2024.

If you use any part of this document, please, refer to
*Mateusz Dolata, Norbert Lange, and Gerhard Schwabe. 2024.* **Development in times of hype: How freelancers explore Generative AI?** *In 2024 IEEE/ACM 46th International Conference on Software Engineering (ICSE '24), April 14–20, 2024, Lisbon, Portugal. ACM, New York, NY, USA, 13 pages.* https://doi.org/10.1145/3597503.3639111

The supplementary material provides a full comparison between challenges reported in SE4AI literature and challenges reported by freelancers participating in the study (see 1.1 below). Additionally, it provides additional information concerning methods applied throughout the study (see 2.1 – 2.3).

# 1 Results / Discussion

## 1.1 Extended Version of Table 2

The table below is an extended version of Table 2 from the manuscript. It includes subcategories which were identified in the literature but were either confirmed by freelancers without further deliberation (↔) or no comments were made by freelancers on the given category of challenges (◇). The star, ★, points to categories for which freelancers identified unique or new challenges. ↗ indicate which factors contribute to upholding challenges regarding GenAI in the category. ↘ informs when a factor lessened the impact of challenges previously reported for AI/ML. ↑ indicates factors that significantly amplify challenges belonging to the given category.

| KA | Subcategory | Challenges and Needs in SE4AI / SE4ML literature [1, 2, 4, 5] | Challenges and Needs Mentioned by Freelancers (FL) | Impact Factors |
|---|---|---|---|---|
| Software Requirements | Customers' Expectations | - unrealistic 100% accuracy expectations,<br>- unrealistic expectations toward adoption (e.g., running system with too little data) | - clients have too high expectations towards abilities of AI because of the hype around GenAI | ↑ fashion |
| | | | - clients demand use of GenAI despite mismatch between the business requirements and GenAI's capabilities | ↑ novelty |
| | Customers' Limited Knowledge | - lack of literacy concerning the capabilities of AI,<br>- lack of knowledge on quantitative metrics to measure those capabilities | - clients use demanding/unrealistic projects to learn about limitations of GenAI generating risks for FL's rating | ↗ novelty |
| | | | - new non-technical clients request GenAI-based solutions due to inflated expectations about GenAI's capabilities | ↗ fashion |
| | Data Limits Requirements | - data limitations go against requirements,<br>- extraction of features from data difficult | *FL do not refer to those and similar problems in relation to Software Requirements* | ◇ |
| | Metrics vs. Requirements | - statistical metrics do not match requirements,<br>- statistical metrics do not match business metrics,<br>- difficult to use requirements coverage method,<br>- lack of coverage-oriented datasets,<br>- no operationalization of coverage for ML | - clients and FL lack effective statistical measures for quality assessment of generated content (as opposed, e.g., to precision and recall measures for classification tasks) | ★ paradigm |
| | | | - assessment of generated content requires domain expertise | ↑ paradigm |
| | | | - reliable ground truth is very difficult / impossible to create so evaluation cannot be reliably automated | ★ paradigm |
| | | | - clients lack adequate business or quality criteria for new tasks if they haven't been previously conducted by humans | ↑ fashion<br>↑ novelty |
| | New Types of Requirements | - fragmented/incomplete knowledge of new or ML-typical non-functional requirements,<br>- requirements hard to elicit / vague / not apparent | *FL confirm the presence of vague and hard-to-elicit requirements* | ↔ |
| Software Design | Components Orchestration | - difficult dependencies between all parts of ML-based system,<br>- additional complexity due to distributed architecture,<br>- hard-to-manage interactions between various ML models, | - orchestration was very difficult after release of some LLMs due to their novelty but lessened with new orchestration frameworks | ↘ novelty |
| | | | - downstream compatibility is hard to achieve and maintain due to nondeterminism of GenAI's output format | ★ paradigm |
| | Inherent Complexity of ML | - changing anything changes everything,<br>- paradigm shift to pipeline-driven / system-wide view,<br>- entanglement created by ML models,<br>- abstraction boundary erosion | - singular LLM's capabilities (e.g., generating answer vs. generating summary) are impossible to separate technically because they rely on a single model | ★ paradigm |
| | | | - similar or seemingly identical prompts can trigger different capabilities based on provision of different context data or subtle differences between prompts | ★ paradigm |
| | | | - outputs of GenAI cannot be meaningfully explained or interpreted (black-box) | ★ paradigm |
| | (Anti-)Patterns | - new anti-patterns due to complexity of ML including glue code, pipeline jungles, correction cascades, dead experimental code paths, technical debt,<br>- lack of good patterns | - anti-patterns including glue code and correction cascades emerge due to nondeterminism of the LLMs' output's format | ↗ paradigm<br>↗ product |
| | | | - good patterns are lacking due to new characteristics of LLMs | ↗ novelty |
| | | | - hype attracts freelancers who reinforce anti-patterns due to lack of development experience | ↗ fashion |
| **Software Construction** | Data Dependency | - dependency on quality and availability of data during training,<br>- continuous validation necessary,<br>- poor calibration of data and models | - dependency on quality and availability of training data lessened due to availability of pre-trained models | ↘ product |
| | | | *FL confirm such problems wrong format, inadequate data, too much or not enough data in context of retrieval augmented generation* | ↔ |
| | Domain Dependency | - reusing models across domains and contexts difficult,<br>- transferring systems across domains,<br>- inability of models to generalize beyond training set | - problems related to transfer, generalizability, and reuse of models lessened due to generality and domain-independency of the available LLMs | ↘ paradigm<br>↘ product |
| | | | - some languages (Arabic) are insufficiently covered by LLMs which was also a problem in earlier NLP solutions | ↔ |
| | Insufficient Knowledge | - lack of understanding of models, libraries, and techniques,<br>- insufficient knowledge about the problem at hand | - shared knowledge base and reference communities to obtain this knowledge are missing for some application scenarios due to novelty | ↑ novelty |
| | Insufficient Tools | - inflexible AI/ML development tools,<br>- fragmented toolchains,<br>- lack of tool infrastructure | *FL did not mention problems related to development tools (such as IDEs).* | ◇ |
| | Integration | - presence of incompatible programming languages,<br>- bugs in libraries, frameworks, and platforms,<br>- interoperability problems | - bugs are present in database platforms (pinecone), orchestration frameworks, APIs due to their novelty | ↗ novelty |
| | | | - frameworks or APIs undergo frequent, unannounced updates to turning developed pipelines ineffective | ↑ product |
| | | | - context window and prompt length are limited in the currently available LLMs and APIs | ↑ product |
| | | | - high latency times slow down the development and reduce user experience due to high load on externally hosted LLMs | ↑ product |
| | Trial and Error | - problems in evaluating and debugging models,<br>- training models requires many iterations,<br>- analyzing and understanding structure and behavior of ML models, | - trial-and-error remains necessary but relates to analyzing existing models through prompting rather than training new models | ↗ paradigm<br>↗ product |
| | | | - GenAI's reliance on billions of parameters makes trial-and-error for training and fine-tuning more costly and time-consuming | ↗ product |
| | | | - need for trial-and-error increased because models more complex and untransparent concerning, e.g., data used for training | ↗ paradigm<br>↗ product |
| | Unreliable Output | - difficult control of quality during development of applications and models,<br>- analyzing and understanding the structure and behavior of the model | - hallucinations occur in LLMs' output | ↑ paradigm |
| | | | - output's format and content are inconsistent | ↑ paradigm |
| | | | - output seems random and unpredictable | ↑ paradigm |
| | | | - output has low quality due to reasoning mistakes | ↑ paradigm |
| | | | - sources used by the model for answer generation are unclear and lack references | ↑ paradigm |
| | | | - debugging is more difficult due to unreliable output | ↑ paradigm |
| **Software Testing** | Experiment Complexity | - difficult to troubleshoot propagated data errors,<br>- inherent variability behind parameters,<br>- difficult to trace results due to long experiments and complex interactions | - inconsistency and variety of results from models make experiments more complex and require more experiments | ↗ paradigm |
| | | | - models' capabilities can change daily (e.g., models might 'unlearn' some capabilities) due to constant training / learning | ↑ product |
| | | | - tracking experiment results is more difficult and not supported by APIs or frameworks | ↑ product |
| | Reproducibility | - hard to reproduce bugs and results due to non-determinism of results,<br>- models' behaviors cannot be completely predicted making validation and verification difficult | - inherent nondeterminism of most GenAI paradigms like LLMs, which can be only partially reduced (e.g., temperature) but never turned off, makes test results hard to reproduce | ↑ paradigm |
| | | | - missing reproducibility enhances where user generated or dynamically collected data is used as context for a prompt due to potential interaction between context and prompt | ↑ paradigm |
| | | | - LLMs exhibit unstable performance with the same task | ↑ paradigm |
| | Testing Criteria | - lack of specification to test against,<br>- discovery and definition of adversarial examples, | - quantitative measures are lacking for generated content due to output's complex character | ↑ paradigm |
| | | | - content and form of output is less predictable than conventional ML due to generative nature | ↑ paradigm |
| | | - real-world testing difficult due to safety or security | - opaqueness of models makes it difficult to specify adequate criteria for validation | ↑ paradigm |
| | Testing Data | - identification of test oracles,<br>- missing access to high-quality test data,<br>- manual labor necessary for creation of test data,<br>- missing test cases | - manual creation of testing data is not feasible because many projects small and at the proof-of-concept stage | ↗ fashion |
| | | | - testing data and test cases is missing due to customers' ad-hoc interest and innovative character of some projects | ↑ fashion |
| | Testing Management | - costs and resources for systematic tests,<br>- management of tests and their results,<br>- unnecessary repetition of tests,<br>- keeping track of tests | *FL do not see testing management as an issue since most of them do not engage in systematic testing.* | ◇ |
| | Testing Standards | - lack of methods for preparation of testing,<br>- major role of developers' assumptions in simulation test | *FL confirm the lack of standards and processes to guide testing, but due to low importance of systematic tests in their projects, it is not a core issue.* | ◇ |
| | Testing Tools and Support | - lack of test environments with trained ML models,<br>- lack of cross-framework and cross-platform support,<br>- lack of tools for testing | - specific tools for testing of GenAI technologies are lacking due to the novelty of the paradigm | ↗ novelty |
| | | | - new testing processes are necessary for GenAI which are yet to come due to novelty of the paradigm | ↗ novelty |
| **Software Quality** | External Dependencies | - hard to understand the effects of external ML algorithms on desired qualities at runtime,<br>- need to deal with not-assured components | - quality assurance of applications, which rely on external, non-assured models, is difficult due to lack of control over core functionality models | ↗ paradigm<br>↗ product |
| | | | - quality assurance is difficult due to black-box, untransparent nature of available LLMs | ↗ paradigm<br>↗ product |
| | New Types of Quality Features (Ethics, Fairness, …) | - new, ML-specific regulatory and ethical consideration of ML models,<br>- lack of established methods to assure fairness, safety, and security of ML | *FL have concerns about the quality features but do not list any new and do not indicate that any of those types amplified due to GenAI.* | ↔ |
| | Quality Standards and Criteria | - lack of criteria for the new types of quality features,<br>- hard to define quality standards,<br>- scalability of quality standards,<br>- lack of certification and qualification, lack of standards on code quality for ML systems | - effective and persistent guardrails for GenAI to assure the quality features are hard to specify due to variable output | ↗ paradigm |
| | | | - desired quality of output and possible limitations imposed by guardrails are hard to balance due to untransparent relation between input and output | ↗ paradigm |
| | | | - sufficient and adequate quality criteria for GenAI are missing (e.g., fairness metrics for LLMs' output) | ↗ paradigm<br>↗ novelty |
| | Quality Assurance Data | - impossible to collect data which is necessary for assuring quality (e.g., individual demographics to assure fairness),<br>- lack of coarse level methods | *FL did not mention any specific challenges related to lack of data for quality assurance.* | ◇ |
| **Software Maintenance** | Adversarial Attacks | - hard to prevent hacking or game rewards function,<br>- risks of extraction of model information by querying,<br>- risks of recovering training set | *FL did not mention purposeful malign behavior and adversarial attacks as major risk for their projects.* | ◇ |
| | Detecting Model Risks | - risk of feedback loops, concept drift, fluctuations in input data, or manipulation,<br>- large scope of tracking and controlling change,<br>- need to avoid damage to environment | *FL did not mention model risks as a significant challenge for themselves, but they see it as a challenge for the providers.* | ◇ |
| | Inadequate Context of Use | - undeclared consumers who use system not in line with original purpose or intentions,<br>- adaptation to changing environment of use,<br>- inadequate user interfaces on top of ML systems | - difficult to predict or limit what input or context will be provided to the system by the user (if application allows for free user input, which most do) due to LLMs acceptability of unstructured input | ↑ product |
| | | | - LLMs can handle unexpected input, but applications might inadequately process LLMs' output based on such input | ★ product |
| | Revalidation of Updated Models | - hard to determine frequency of training due to changes in context,<br>- need to revalidate after updates, | - revalidation is hard due to frequent changes in the models delivered by external parties (e.g., OpenAI) | ↑ product |
| | | | - downstream compatibility is difficult to assure due to changes in the models and external dependencies | ↑ product |
| | | - updates in ML consider code, model, data as opposed to code in legacy systems, | - applications' usability or usefulness difficult to guarantee over time as no information available on plans of the providers | ↑ product |
| | | - changes in model behavior negatively impact users' trust and downstream/backward compatibility | - new requirements for validation since validation is necessary of the applications and prompts used therein and not the models themselves | ★ product |
| | Tracking Data | - keeping track of datasets and their metadata | *FL did not refer to challenges related to tracking data.* | ◇ |
| Software Configuration Management | Infrastructure | - technical debts present in ML frameworks, <br> - need for reliable infrastructure to run the model | *FL mention need for infrastructure if they need to host data for retrieval augmented generation or if they want to host their own model (two cases in the data), however those problems are not considered new or amplified.* | ↔ |
| | Economic and Computing Resources | - timing, memory, and energy constraints, <br> - costs of training, <br> - long training times for iterations, <br> - need for adequate computing power | - new cost structure is necessary due to pay-per-use policy of major providers including costs of exploring external models' potentials and costs of using external models in the applications | ★ product |
| | | | - resources necessary for training of own GenAI models are incomparable to conv. ML due to models' complexity and size | ★ paradigm |
| | Scalability | - need to scale models at the end of production, <br> - need for scalable oversight for models | *FL refer to speed and workload issues of external models, indicating that this might be related to their scalability. This negatively impacts the scalability of FL's own applications, but it is similar to earlier challenges.* | ↔ |
| | Short-Lived Technology | - AI-based system can be invalidated due to trend changes | *FL acknowledge that GenAI is subject to the hype, but do not expect that the trend will disappear quickly.* | ↔ |
| | Third-Party Dependency | *even though external dependencies are considered for quality assurance or development, considered literature does not explicitly attend to implications of external dependencies for software configuration management* | - dependency on external providers is significantly higher because applications rely on external models, access to APIs, and providers' infrastructure | ★ product |
| | | | - the usability and popularity of FL's applications depends on speed, availability, and compatibility of external components | ★ product |
| | | | - planning of economic resources is difficult due unclear plans and pricing strategies of external providers | ★ product |
| SE Process, Models & Methods | Engineering | - need for continuous engineering, <br> - necessity for ad-hoc development | - engineering is ad-hoc and results from short-term demand driven, e.g., by hype around GenAI | ↗ fashion |
| | Lack of Adequate Process Models | - need for a highly iterative life-cycle model, <br> - lack of guidance on various aspects, <br> - lack of well-defined, widely accepted process model, <br> - lack of processes for annotation | *FL did not notice specific challenges concerning SE process models as they have engaged in ad-hoc development before. However, they agree that they could benefit from more precise, systematic guidance on how to deal with the new paradigm of GenAI.* | ↔ |
| SE Professional Practice | Customer Communication | - hard to negotiate unfeasible expectations, <br> - hard to convince clients to pay continuously for improvements, <br> - need for skills to help customers set feasible targets / requirements | - common ground with clients from outside IT context is difficult due to hype and inflated expectations around GenAI | ↗ fashion |
| | | | - communication with clients is difficult because clients rely on misinformation and inflated promises regarding GenAI | ↗ fashion |
| | End-user Engagement | - clients focus on technical solutions rather than involving end-users, <br> - skepticism of users who were involved not early enough | - clients without experience in IT projects refuse to participate in requirements elicitation or provide requirements which hare not useful | ↗ fashion |
| | | | - hype-driven clients expect fast results neglecting the need to involve end users in design | ↗ fashion |
| | Explanations | - difficult to explain to clients why models produces certain output, <br> - insufficient explainability for some model types, <br> - explaining the potential for improvement over time | - performance of LLMs and GenAI models difficult to explain to clients due to lack of knowledge about how models work | ↑ paradigm |
| | | | - opaqueness and inherent variability of the models make it difficult to understand output and make reliable statements | ↑ paradigm |
| | | | - clients blame freelancers for low performance of the models | ↑ paradigm |
| | Societal Concerns | - need for continuous chain of human responsibility, <br> - ethical considerations, | *FL acknowledge that AI technologies have societal implications and try to implement guiderails to prevent them (see SW Quality).* | ↔ |
| | | - proliferation of fake content based on generation,<br>- speed of technical progress higher than regulation | *FL see the risk of job losses in the society, yet do not consider it a challenge for their own development activity.* | ↔ |
| | Team Composition and Collaboration | - variety of disciplines and backgrounds involved in AI projects,<br>- overlaps between data scientists and engineers,<br>- slow start to onboard non-engineers | *FL work alone or in small teams of FL (when a freelancing agency) and did not mention problems concerning team composition or collaboration.* | ◇ |
| | Workforce Skills | - need for diverse skills,<br>- lack of diverse skills in teams,<br>- lack of adequately educated engineers | - skilled freelancers are lacking due to increased hype-driven demand for their skills and experience | ↑ fashion |
| | | | - freelancers without adequate experience join the market due to enhanced demand and potential for lucrative jobs | ↑ fashion |
| | | | - staying on top of dynamic market changes and technical developments causes major effort to keep pace with it | ↑ fashion |
| | Community and Competition | *the literature does not explicitly mention community-related topics as challenges* | - community to exchange with is difficult to find, especially on use of LLMs and GenAI for specific domains, due to novelty and niche character of some projects | ★ novelty |
| | | | - competition increases due to many new freelancers entering the market | ★ fashion |
| SE Management | Planning Uncertainty | - uncertainty of estimating development time and cost,<br>- no well-defined input-output factor for validation,<br>- assessment of long-term potential based on short-term metrics,<br>- impossibility to make guarantees on cost-effectiveness | - estimations are difficult to make due to lack of experience | ↑ novelty |
| | | | - limited ability to say upfront if a specific case can be covered by GenAI technologies due to lack of experience | ↑ novelty |
| | | | - clients expect that a certain task will be solved with GenAI despite more adequate and easier-to-plan alternatives, e.g., conventional ML | ↗ fashion<br>↗ paradigm |
| | | | - reasonable offer / putting a price tag on a project difficult due to lack of heuristics and experience | ↑ novelty |
| | Regulatory Demands | - amount of reviews, software updates, cycles of data collection or annotation demanded by law | *FL did not refer to regulatory aspect as a challenge for them.* | ◇ |
| | Resources Shortage and Costs | - resource limits curbing down efforts,<br>- lack of organizational incentives,<br>- lack of resources | - costs related to experimentation during prompt-engineering are hard to manage and control especially if the client does not cover those costs | ↗ product |
| | Assessment of Gains | - quantification and assessment of the knowledge generated during the development of the system | *FL did not refer to this aspect as a challenge for them, they mention that freelancing allows for learning, and they see each project as learning opportunity.* | ↔ |
| | Pre-Selection Effort | *the literature does not explicitly mention pre-selection as challenges* | - preselecting projects is hard based on the requests from clients due to lack of experience | ★ novelty |
| | | | - fashionableness causes many unusual or incomplete requests and assessing them requires effort to explore whether GenAI technologies are useful to solve the requests | ★ fashion |
| | | | - frequent changes and instability of the models might invalidate preselection decisions | ★ product |
| Training & Testing Data | Data Access | - hard to discover what data exists and where it is,<br>- difficulty to collect new data and prepare it | *FL refer to challenges related to data if their projects involve (a) training of fine-tuning models or (b) retrieval augmented generation; in both cases the challenges confirm many of the challenges listed for conventional ML, larger datasets are necessary for training a generative model.* | ↗ paradigm |
| | Data Management | - difficult controlling, versioning, deploying of data | | |
| | Data Preparation | - need to combine, transfer, clean, and transform data,<br>- lack of tools for data preparation,<br>- inconsistent data types and quality | | |
| | Data Quality | - difficult to assure data quality,<br>- missing values in data,<br>- missing rare cases in data,<br>- lack of tools to annotate and assure data quality,<br> - bias in data | - no need for training data and training data quality assurance due to the availability of pre-trained models | ↘ product |

# 2 Method

## 2.1 Participants Acquisition

### 2.1.1 Job Description Posted to Upwork

*I am looking for developers who create software applications involving intelligent features based on generative AI (including any of the recent large language models): it might involve web apps, mobile apps, plug-ins or packages towards bigger projects. I want to learn about problems and challenges you encounter in your daily work with GPT, LLaMA, LaMDA, and others, their APIs, or your own generative models, how you solve those challenges, and what would you wish for in the future.*

*I do not need any sensitive information (employer, name of the project/app, country, private addresses, etc.), but some basic information about the functionality might be necessary for me to understand the use case.*

*These are the selection criteria for interview participants:*
*- software developer or software project manager with 2 years of experience or more,*
*- participated in 2 or more development projects involving the use of generative AI platforms / APIs,*
*- the developed software uses, processes, or interprets the output from the API (i.e., at least one of your projects/apps should be more than simply an adapted/extended version of the OpenAI's playground).*

*Your tasks involve:*
*- filling out a short survey (5-10 min) prior to the interview including an informed consent for participation,*
*- participating in an online interview via zoom (approx. 45-50 min).*

*Each participant will be paid 35 USD after the completion of the interview.*

*The dates for the interview will be arranged individually. The interviews will be audio-recorded. Later on, a research team will anonymize, transcribe, and analyse them.*

*If you think you're the right participant for the interview, drop me a line.*

### 2.1.2 Question Asked in the Proposal Phase

1. *What generative AI platforms or APIs did you use in your past projects?*
2. *How many years of experience as software developer do you have?*
3. *What is your highest training / education degree?*
4. *How many from your recent projects used a generative AI platform or API?*

### 2.1.3 Conflict of Interests Declaration

*The authors, unaffiliated with Upwork Inc., did not receive any benefits from the company or its subsidiaries at any time. The choice was based on specific criteria: (1) the ability to post a job with a fixed rate, to ensure equal pay for all freelancers, (2) detailed search criteria, including GenAI-related topics, (3) a long-standing presence with experienced freelancers, (4) a clear, consistent contract price, to fit University's study participant reimbursement rules. We consulted multiple online resources, e.g., arc.dev/employer-blog/toptal-upwork-fiverr-arc/, www.g2.com/categories/freelance-platforms?attributes[221]=420#grid*

### 2.1.4 Participants List

This list presents all participants of the study with their random codes, demographic data, and data related to developer experience and specific experience with GenAI models / APIs.

| Code * - quotes used in paper | Age (years) | Number of GenAI Projects | Developer Experience (years) | Highest Degree | Education Subject | Country of Residence | GPT | DALL-E | StableDiffusion | Midjourney | LLaMA | LaMDA/Bard | Vicuna | Others |
|---|---|---|---|---|---|---|---|---|---|---|---|---|---|---|
| aa | 25 | 3 | 3 | HND | Website Development | Nigeria | X | | | | | | | |
| ab* | 38 | 5 | 2 | Master | Economics | Serbia | X | | | | X | | X | |
| ac | 30 | 5 | 3 | Bachelor | Material Engineering | Egypt | X | X | X | X | X | X | | |
| ad* | 30 | 4 | 10 | Master | Machine Learning | United States | X | | | | | X | | X |
| ae | 30 | 2 | 4 | Bachelor | History | Montenegro | X | | | | X | | X | |
| af | 34 | 4 | 2 | Bachelor | Business Administration | Germany | X | | X | | | | | |
| ag | 23 | 5 | 5 | Bachelor | Software Engineering | India | X | X | | | | | | |
| ah | 30 | 4 | 6 | Bachelor | Electronic Engineering | India | X | X | X | | X | | | |
| ai | 33 | 4 | 5 | Master | Software Engineering | United States | X | X | | X | | | | X |
| bj | 24 | 7 | 3 | Master | Computer Science | Serbia | X | | | | X | | X | |
| bk | 24 | 7 | 4 | Master | Computer Science | India | X | | X | X | | | | |
| cl* | 55 | 2 | 15 | Bachelor | Computer Science | United States | X | | X | | | | | X |
| cm | 40 | 6 | 15 | Master | Software Engineering | United States | X | X | X | | | | | |
| dn | 23 | 4 | 5 | Bachelor | Geography | Nigeria | X | X | X | | X | | | |
| do* | 31 | 5 | 5 | Master | Computer Science | Kenya | X | X | | X | | | | |
| dp | 22 | 3 | 5 | Student | Computer Science | Poland | X | X | | | | | | |
| er | 43 | 3 | 7 | Bachelor | Computer Science | Turkey | X | X | X | X | | | | X |
| es | 37 | 4 | 10 | Bachelor | Electronic Engineering | Indonesia | X | X | | | | | | X |
| et | 38 | 7 | 15 | Bachelor | Computer Science | United States | X | | X | | | | | |
| eu* | 35 | 4 | 15 | Master | IT Management | Serbia | X | | | | X | | X | |
| ev* | 38 | 6 | 13 | Master | Computer Science | France | X | | X | | | | | X |
| gw | 36 | 11 | 10 | PhD | Machine Learning | Netherlands | X | X | | | | | | |
| gx | 39 | 7 | 20 | Master | Human Computer Interaction | United States | X | | X | | X | | | |
| hy | 29 | 5 | 5 | Bachelor | Computer Science | Pakistan | X | | | | X | | | |
| hz | 35 | 3 | 10 | Master | Software Engineering | France | X | | | | | | | |
| ia* | 28 | 3 | 5 | PhD | Physics | Romania | X | | X | | | | | |
| jb | 31 | 4 | 7 | Bachelor | Computer Science | Portugal | X | X | X | | | | | |
| jc | 24 | 5 | 2 | Bachelor | Information Systems | United States | X | | | | | | | X |
| jd | 42 | 5 | 20 | Bachelor | Computer Science | United Kingdom | X | | X | X | | X | X | |
| je | 42 | 3 | 10 | PhD | Physics | United Kingdom | X | | | | | | | |
| jf* | 23 | 5 | 5 | Bachelor | Computer Science | Romania | X | | | | | | | |
| kg | 33 | 10 | 10 | Master | Physics | Sweden | X | X | X | X | | | | X |
| kh | 32 | 5 | 7 | Bachelor | Computer Science | India | X | X | | | | | | |
| ki | 30 | 2 | 1 | Bachelor | Electronic Engineering | Sri Lanka | X | | | | | | | |
| lj | 52 | 6 | 10 | PhD | Telecommunications | France | X | | | | | | | |
| mk | 35 | 2 | 3 | Master | Communication | New Zealand | X | | | | | | | |
| ml* | 30 | 4 | 5 | PhD | Machine Learning | Pakistan | X | | X | | X | X | | X |
| om | 23 | 5 | 3 | Student | Geography | Nigeria | X | X | X | X | | | | |
| pn | 38 | 2 | 2 | Master | Telecommunications | Montenegro | X | | | | | | | |
| ro | 36 | 15 | 6 | Master | Machine Learning | India | X | X | | | X | X | | X |
| rp | 31 | 3 | 3 | Bachelor | Commerce | Dominican Republic | X | | X | X | | | | |
| sr | 25 | 4 | 2 | Bachelor | Computer Science | Pakistan | X | | | | | | | |
| ss* | 30 | 5 | 5 | Master | Mathematics | Pakistan | X | X | X | | | | | |
| st | 23 | 6 | 3 | Bachelor | Computer Science | Nigeria | X | X | X | X | | | | |
| su | 34 | 4 | 9 | Master | Computer Science | Serbia | X | X | X | X | X | | | |
| tv | 24 | 5 | 7 | Master | Computer Science | Pakistan | X | X | | | | | | |
| tw | 27 | 3 | 2 | Bachelor | Economics | Germany | X | | | | | | | |
| tx* | 30 | 3 | 6 | Master | Control & Automation Engineering | Turkey | X | X | | | | | | |
| ty | 51 | 8 | 15 | PhD | Natural language processing | Finland | X | | | | | | | |
| uz | 30 | 5 | 2 | Master | Electronic Engineering | Pakistan | X | X | X | | | | | X |
| va* | 19 | 3 | 3 | Student | Computer Science | Ukraine | X | X | X | | | | | X |
| wb | 49 | 3 | 8 | Master | Psychology | Portugal | X | | | | | | | |
| SUM | | | | | | | 52 | 21 | 18 | 17 | 13 | 6 | 5 | 11 |
| AVG | 32.6 | 4.8 | 6.9 | | | | | | | | | | | |
| MED | 31 | 4 | 5 | | | | | | | | | | | |

### 2.1.5 Quoted Participants

This list provides more information about the participants quoted in the manuscript.

| Interviewee Code | Age (years) | No. GenAI Projects | Experience (years) | Highest Degree | Education Subject (CS – Computer Science, ML – Machine Learning, CA – Control. & Autom. Eng.) | Country of residence | Declared professional experience concerning GenAI platforms (DE – DALL-E, MJ – Midjourney, SD – Stable Diffusion, custom – custom self-trained models) |
|---|---|---|---|---|---|---|---|
| - Two exemplary use cases for which the interviewee developed solutions ||||||||
| ab | 38 | 5 | 2 | MSc | Economics | Serbia | GPT, LLaMA, Vicuna |
| - Classifying SaaS websites with GPT and Vicuna<br>- Investment and trading prediction using GPT and LLaMA ||||||||
| ad | 30 | 4 | 10 | MSc | ML / AI | USA | GPT, LaMDA, Others |
| - AI sales assistant based on fine-tuned GPT<br>- Recommendation chatbot for retail based on fine-tuned GPT ||||||||
| cl | 55 | 2 | 15 | BSc | CS | USA | GPT, MJ, PaLM, SD |
| - Generative shoe design with text-to-image models<br>- Designing Kubernetes cloud environment using GPT and Bard/LaMDA ||||||||
| do | 31 | 5 | 5 | MSc | CS | Kenya | DE, GPT, LLaMA |
| - Text summarization of demand letters in financial industry using GPT<br>- API for Q&A bots for text-to-text and table-to-text scnearios ||||||||
| eu | 35 | 4 | 15 | MSc | IT Mngmt. | Serbia | GPT, LLaMA, Vicuna |
| - Q&A chatbot for local knowledge base based on GPT<br>- Collection of news on a topic and providing summary – GPT ||||||||
| ev | 38 | 6 | 13 | MSc | CS | France | GPT, MJ, Others |
| - Customer support chatbot for home appliances based on GPT<br>- Integrating Midjourney output in an AR application in luxury industry ||||||||
| ia | 28 | 3 | 5 | PhD | Physics | Romania | GPT, MJ |
| - Summarization of market research reports based on GPT<br>- Summarization of output obtained from an SQL database using GPT ||||||||
| jf | 23 | 5 | 5 | BSc | CS | Romania | GPT |
| - Customer support assistant using dynamically scrapped content, uses GPT<br>- Q&A chatbot as a medical assistant based on GPT ||||||||
| ml | 30 | 4 | 5 | PhD | ML / AI | Pakistan | GPT, LLaMA, SD, LaMDA |
| - Customized medical prescription recommendation based on GPT<br>- Modules for search and Q&A for regulatory documents based on GPT ||||||||
| ss | 30 | 5 | 5 | MSc | Math. | Pakistan | DE, GPT, SD, custom |
| - Generation of videos with avatars reading input text, with various models<br>- Mobile guidance app using video input and output with custom model ||||||||
| tx | 30 | 3 | 6 | MSc | CA | Turkey | DE, GPT |
| - Generation of Game Design Documents for mobile games based on GPT<br>- Instant game content generation by player with voice based based on GPT ||||||||
| va | 19 | 3 | 3 | Stud. | CS | Ukraine | DE, GPT, SD, Others |
| - Teaching assistant Q&A-bot providing sources, images, and summaries<br>- Generating a multi-perspective, multi-opinion newsletter with GPT ||||||||

## 2.2 Data Collection

### 2.2.1 Survey Questions

Survey was hosted on a LimeSurvey server of the university. Participants received invitation to the survey after accepting the contract on Upwork. All communication with the participants was happening through Upwork.

[Informed Consent]

1. Which of the following tools did you use in the apps you developed?
    - AlphaFold
    - BLOOM
    - Chinchilla AI
    - DALL-E
    - GPT / ChatGPT
    - LaMDA / Bard
    - LLaMA
    - Midjourney
    - PaLM
    - Stable Diffusion
    - Other: [...]

2. In what projects and for what did you use the tools? (open question)
   *Provide two or more examples. A brief description is enough, e.g., "Connected to GPT to create ideas on a given topic in a brainstorming app." or "Used Stable Diffusion to generate graphical summaries of daily news for a news app." If you want, feel free to paste external links.*
   *Please, give preference to projects or usage scenarios which challenged you most.*
3. What is the biggest challenge when creating apps with generative AI capabilities? (open question)
4. What is the biggest reward from integrating generative AI capabilities in apps? (open question)

### 2.2.2 Interview Guide

After the survey, participants were provided access to a calendar in which they could book an appointment that suits their time schedule. After booking the appointment, they received access to the interview guide so they could see it before the interview (but they were not requested to do so).

Interview "Developing Apps that use Generative AI"

The interview will be conducted in a semi-structured manner. It means, I will try to address most of the topics listed below. However, we might change the order of the questions depending on how the conversation develops. Some of the questions might not perfectly fit your specific situation or experience. Do not worry. We will figure out how to go about it.

Part 1: Opening
Short Introduction
- *Goal of the study*
- *My background*
- *Your background*

Part 2: Example Projects
Which project based on generative AI are you most proud of? Why?
Were there any projects involving generative AI that did not work out so well? (optional)
Is there any other project you want to tell me about? (optional)

Part 3: Comparison to Projects which don't Involve Generative AI

*What has surprised you most about projects involving generative AI so far?*
*How are generative AI projects different from / similar to earlier projects?*
*What is the biggest uncertainty in generative AI projects? How do you deal with them?*

*Part 4: General Attitude*
*In the survey you mentioned the following challenge: …  (for each challenge listed in the survey)*
*Can you explain to me what exactly you mean by this? Why is this a challenge?*
*What does make the most fun to you when you are developing software based on generative AI?*
*What special skills does a person need to develop software based on generative AI?*
*What are the positive and negative implications of generative AI for you as the freelancer?*

## 2.3 Data Analysis

### 2.3.1 Interviews – Initial Coding
(coder: 2nd author, approach: bottom up, goal: identify passages pointing to challenges)

In the early phase of analysis, the coder decided to identify challenges based on four different sensitizing perspectives: ethics, society, emotions, and technology inspired by the coders' professional background (independent consultant and coach for teams and individuals in IT industry). The coding was supervised by both other authors and edge cases were discussed among them to assure reliability. We provide those codes here for transparency, but we want to stress that this differentiation was not further used during the analysis. Instead, the authors regrouped all 268 segments as described below. Here are the four categories used in the first round of coding along with the number of segments that were subsumed under those codes (numbers do not sum up to 268 as some passages received more than one code):
- ethical challenges (21 segments),
- social challenges (23 segments),
- emotional challenges (86 segments),
- technical challenges (141 segments).
- OVERALL: 268 segments including challenges from the freelancers' perspective.

### 2.3.2 Interviews – Workshops and Sorting according to SWEBOK

After the identification of 268 segments describing freelancers' perspective on the challenges related to GenAI, authors conducted two interpretation workshops of two hours each in which they discussed the meaning of the challenges and potential ways to make sense of them. They were provided the selected passages prior to the first workshop to become familiar with the content of the data. The workshops were moderated by the first author and involved various deductive and inductive sorting approaches. Finally, inspired by the relevant literature in the field, authors settled on classification of the segments according to SWEBOK Knowledge Areas (extended by the KA Training and Testing Data) to reflect approaches implemented in meta-reviews, which assured best comparability to existing literature. We used the official SWEBOK descriptions and the exsiting categorizations from existing meta-reviews [1, 2] to guide the categorization conducted collaboratively.

| Knowledge Area (number of coded segments) | Description (based on SWEBOK) | Example |
|---|---|---|
| Software Requirements (41) | The Software Requirements KA is concerned with the elicitation, negotiation, analysis, specification, and validation of software requirements. | *Yes, I think I think the main problem is really, for example, when when you talk with a client about the success criteria, they expect, for example, to have quantified accuracy metrics of that specific use case. (ad)* |
| Software Design (31) | Design is defined as both the process of defining the architecture, components, interfaces, and other characteristics of a system or component and the result of that process. | *I think the challenge is understanding the LangChain and how it works. It's important to know that you just can't dump all the data and expect, you know, a large language model or the app to extract it. It needs to be converted. The data needs to be first converted into small size chunks. Understanding chunk overlap is an important concept. (…) Then converting those text into embeddings, it's an important concept. (…) You cannot store text as it is in a SQL database. (…) There are a lot of new technologies that need to be that you need to really clear about before building an app. (ai)* |

| Software Construction (63) | This KA covers software construction fundamentals; managing software construction; construction technologies; practical considerations; and software construction tools. | *The biggest challenge is getting the right prompt to use, because sometimes even after you finish, you think you are done working. Then you input the API keys or the prompts at the end of the day. It doesn't work that you have to like keep trying and keep trying until you get the perfect prompts you want. (dn)* |
|---|---|---|
| Software Testing (12) | The Software Testing KA includes the fundamentals of software testing; testing techniques; human-computer user interface testing and evaluation; test-related measures; and practical considerations. | *You cannot simply build a predictor for X, and then the customer can just say: 'you are 10% off the goal, so, it's not working, sorry'. With generative AI, it's so subjective and it is very difficult to have a quantitative evaluation for a project and it's hard to to understand for the customers as well how to how to do that. (gw)* |
| Software Quality (12) | Software Quality KA includes fundamentals of software quality; software quality management processes; and practical considerations. | *I think the feature that we are implementing will depend on the directive in the world. I mean regulation. (…) I think these technologies improve very fast. But for many countries, this regulation can't be produced in that top speed. (…) So, I think we should think about that regulation. (ki)* |
| Software Maintenance (14) | The Software Maintenance KA includes fundamentals of software maintenance; key issues in software maintenance; the maintenance process; software maintenance techniques; disaster recovery techniques, and software maintenance tools. | *I would say that I am really a bit anxious about it, because like tomorrow OpenAI will change something and from that moment on [the app] will not work. (…) I could say 'Client, that's okay, you are on your own. I don't care about what will happen afterwards'. My policy is that I give guarantee that if I miss something, I will be back. I will fix it because it's my bad. (va)* |
| Software Configuration Management (25) | The Software Configuration Management KA covers management of the SCM process; software configuration identification, control, status accounting, and auditing; software release management and delivery; and software configuration management tools. | *So when we're talking about GPT, it's a lot of people want to fine tune the model (…) Because if you do understand what it means, then you know, the job that was $500 is now going to be $5,000 because you have to run an instance. You have to run a whole backend server and all of those things. You can't just make use of APIs and things like that anymore. (rp)* |
| Software Engineering Process, Models and Methods (3) | Topics covered include process implementation and change; process definition; process assessment models and methods; measurement; and software process tools. Further, they also involve modeling; types of models; analysis; and software development methods. | *Firstly, we have to define our requirement what kind of output we want to extract. Once we define our requirement, we can use OpenAI playground to try. Once, we get our output, we can let it go. Once we get at a better point, we can apply it with OpenAI API. So that is the process. (ki)* |
| Software Engineering Professional Practice (41) | The Software Engineering Professional Practice KA covers professionalism; codes of ethics; group dynamics; and communication skills. | *So, I would say that the biggest challenge was that it's a new industry and it's developed very fast. And some articles that you read today, are not working anymore tomorrow. So yeah, it's changing very, very fast and you don't really know like what will work and what will not until you try. And they are not that much information in general because there are no such stable communities. (va)* |
| Software Engineering Management (14) | The Software Engineering Management KA covers initiation and scope definition; software project planning; software project enactment; product acceptance; review and analysis of project performance; project closure; and software management tools. | *I'm definitely safer in this approach where I have to be very I mean, I have to be very sure if I take on the project, so I can deliver great results for the client, and that client will be happy with the results. I think this definitely has changed, I mean, the fact that I take on less projects regarding generative AI, because I like to be, I like to be comfortable. (dp)* |
| Training and Testing Data (12) | We derived this KA from earlier meta-reviews concerning SE4AI / SE4ML. It includes challenges related to generation, management, and use of training and test datasets for ML. | *The quality of data varies from plan to plan and from project to project based on what data they have and what they are trying to make of that data. And many of the times the model itself is able to handle some bad quality data like some missing data. The model itself are able to handle that. (ag)* |

One of the intermediate results of the workshops was the identification of factors that were claimed to contribute to the challenges, as the interviewees were explaining where a challenge was coming from. The four factors were: hype – fashionableness of the GenAI, hype – novelty of the GenAI, technology – characteristics of GenAI paradigm, technology – characteristics of GenAI product. We used those four categories as an interpretative sensemaking frame to establish a shared understanding between the authors according to interpretative research approach [3, 6, 7] and used them to structure the results.

References

[1] Giray, G. 2021. A software engineering perspective on engineering machine learning systems: State of the art and challenges. *Journal of Systems and Software*. 180, (Oct. 2021), 111031.

[2] Martínez-Fernández, S., Bogner, J., Franch, X., Oriol, M., Siebert, J., Trendowicz, A., Vollmer, A.M. and Wagner, S. 2022. Software Engineering for AI-Based Systems: A Survey. *ACM Transactions on Software Engineering and Methodology*. 31, 2 (Apr. 2022), 37e:1-37e:59.

[3] McKenzie, P.J. 2005. Interpretative Repertoires. *Theories of information behavior: A researcher's guide*. K.E. Fisher, S. Erdelez, and L. McKechnie, eds. Information Today.

[4] Nahar, N., Zhang, H., Lewis, G., Zhou, S. and Kästner, C. 2023. A Meta-Summary of Challenges in Building Products with ML Components -- Collecting Experiences from 4758+ Practitioners. *Proc. Intl. Conf. on AI Engineering - Software Engineering for AI* (Melbourne, Australia, Mar. 2023).

[5] Paleyes, A., Urma, R.-G. and Lawrence, N.D. 2022. Challenges in Deploying Machine Learning: A Survey of Case Studies. *ACM Computing Surveys*. 55, 6 (Dezember 2022), 114:1-114:29.

[6] Schwartz-Shea, P. and Yanow, D. 2012. *Interpretive research design: concepts and processes*. Routledge.

[7] Stebbins, R.A. 2001. *Exploratory research in the social sciences*. Sage Publications.